\documentclass[twocolumn]{aastex61}
\usepackage{hyperref}

\shorttitle{Probabilistic Constraints on Convective Core Overshooting}
\shortauthors{Rosenfield et al.}

\newcommand{\zsun}{\mbox{$Z_{\odot}$}}
\newcommand{\Msun}{\mbox{$M_{\odot}$}}
\newcommand{\msun}{\mbox{$M_{\odot}$}}

\newcommand{\lambdac}{\mbox{$\Lambda_{\rm c}$}}
\newcommand{\Lambdac}{\mbox{$\Lambda_{\rm c}$}}
\newcommand{\aov}{\mbox{$\alpha_{ov}$}}

\newcommand{\hp}{\mbox{$H_p$}}
\newcommand{\feh}{\mbox{[Fe/H]}}
\newcommand{\asteca}{{\tt ASteCA}}
\newcommand{\ASteCA}{{\tt ASteCA}}
\newcommand{\MATCH}{{\tt MATCH}}
\newcommand{\match}{{\tt MATCH}}

\begin{document}

\title{A New Approach to Convective Core Overshooting: Probabilistic Constraints from Color-Magnitude Diagrams of LMC Clusters}

\correspondingauthor{Philip Rosenfield}
\author{Philip Rosenfield}
\altaffiliation{NSF Astronomy and Astrophysics Postdoctoral Fellow}
\affil{Harvard-Smithsonian Center for Astrophysics}
\email{philip.rosenfield@cfa.harvard.edu}

\author{L\'eo Girardi}
\affil{INAF Padova, Padova, Italy.}

\author{Benjamin F. Williams}
\affil{University of Washington, Seattle, WA, United States.}

\author{L. Clifton Johnson}
\affil{University of California San Diego, San Diego, CA, United States.}

\author{Andrew Dolphin}
\affil{Raytheon Company}

\author{Alessandro Bressan}
\affil{SISSA, Trieste, Italy.}

\author{Daniel Weisz}
\affil{University of California Berkeley, Berkeley, CA}

\author{Julianne J. Dalcanton}
\affil{University of Washington, Seattle, WA, United States.}

\author{Morgan Fouesneau}
\affil{MPIA Heidelberg, Heidelberg, Germany.}

\author{Jason Kalirai}
\affil{Space Telescope Science Institute, Baltimore, MD 21218, USA}

\begin{abstract}
We present a framework to simultaneously constrain the values and uncertainties of the strength of convective  core overshooting, metallicity, extinction,  distance, and age in stellar populations. We then apply the framework to archival Hubble Space Telescope observations of six stellar clusters in the Large Magellanic Cloud that have reported ages between $\sim1-2.5$ Gyr.  Assuming a canonical value of the strength of core convective overshooting, we recover the well-known age-metallicity correlation, and additional correlations between metallicity and extinction and metallicity and distance. If we allow the strength of core overshooting to vary, we find that for intermediate-aged stellar clusters, the measured values of distance and extinction are negligibly effected by uncertainties of core overshooting strength. However, cluster age and metallicity may have disconcertingly large systematic shifts when \lambdac\ is allowed to vary by more than $\pm\ 0.05\  \hp$. Using the six stellar clusters, we combine their posterior distribution functions to obtain the most probable core overshooting value, $0.500^{+0.016}_{-0.134} \hp$, which is in line with canonical values.
\end{abstract}


\section{Introduction}

Stellar evolution models are fundamental to nearly all studies in astrophysics. They play an important role in understanding the initial mass function (IMF) \citep[e.g.,][]{Chabrier2003}, in determining line-of-sight extinction \citep[e.g.,][]{Schlafly2011}, in measuring distances (e.g., via brightness of the tip of the red giant branch) \citep[e.g.,][]{Salaris1997}, in deriving supernovae rates and progenitor masses \citep[e.g.,][]{Smartt2015}, and in measuring the cosmic star formation history \citep[e.g.,][]{Madau2014}. Unfortunately, some important aspects of stellar evolution remain poorly constrained and can impact the interpretation of galaxy observations \citep[e.g,][]{McQuinn2010, Melbourne2012}. These aspects, such as mixing due to rotation or convection, are too complex to be derived from first principles and can only be constrained by observations.

The strongest observational constraints on stellar evolution models come from resolving individual stars in stellar clusters \citep[e.g.,][]{Gallart2005}. Star clusters are excellent stellar physics laboratories because individually, they fill a narrow parameter space in metallicity, abundance, and age, allowing the calibration of aspects of physical models. within the Galaxy, stellar model constraints benefit from precise measurements of surface quantities and abundances of many member stars, and in some cases the possibility of independent measurement techniques from asteroseismology, and reliable parallaxes \citep[e.g.,][]{Torres2010, Overbeek2017}. However, there are not many nearby clusters that are both easily observable and young or intermediate-aged. Nearby Galactic clusters also tend to have near-Solar metallicities, and derived model constraints must then be extrapolated for use in stellar populations elsewhere. This limitation can be partially ameliorated by studying extragalactic star clusters. The Large and Small Magellanic Clouds (LMC, SMC) contain resolvable stellar clusters that are useful for accessing sub-Solar metallicities typical of nearby dwarf galaxies and galaxies in the distant universe. The MCs provide a rich sample of stellar clusters over a broad range in cluster mass and age. The MCs are also close enough to resolve stellar cluster members several magnitudes below the main sequence turn off (MSTO), either using ground-based telescopes for more massive clusters or using the Hubble Space Telescope (HST) for clusters in the denser regions of the MCs.

\begin{figure}
 \includegraphics[width=\columnwidth]{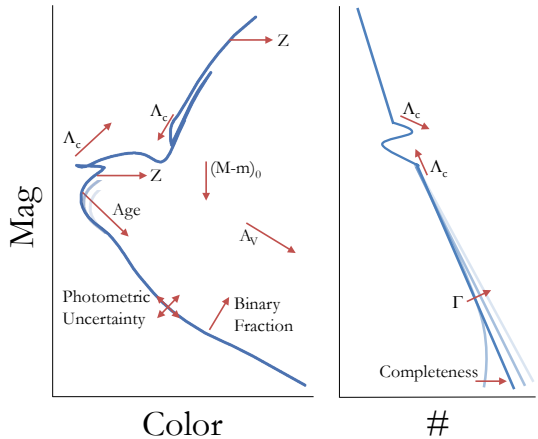}
  \caption{Schematic optical color-magnitude diagram (CMD; left) and luminosity function (right) showing how observational and theoretical uncertain parameters can change the morphology of an intermediate age isochrone or synthetic stellar population drawn from isochrones and an IMF. Each arrow direction denotes approximate change with an increase in that parameter. An increase in photometric uncertainty and binary fraction will spread stars location on the CMD, while the other shown parameters will shift the location of the isochrone. A probabilistic approach is needed to disentangle these effects.}
  \label{fig:cmdlfschem}
\end{figure}

To assess stellar models, researchers fit isochrones or synthetic stellar populations to their observations \citep[e.g,][]{Milone2009,Girardi2009,Goudfrooij2011}. Unfortunately, uncertainties from both models and observations are not always accounted for, and seldom are degeneracies between sources of uncertainty modeled or discussed \citep[with the strong exception of the robust Bayesian analyses led by][]{vonHippel2006}. Figure \ref{fig:cmdlfschem} shows a schematic of how theoretical and observational parameters can shift (1) the morphology of an isochrone on an optical color magnitude diagram (CMD) and (2) the (number density of a) luminosity function of a single intermediate-aged ($\sim1.5$~Gyr) stellar population. Certain combinations of parameters, for example, distance modulus, $\mu_0$, and extinction, $A_V$, could be construed as different age and metallicity, $Z$, of the cluster. One can choose other combinations of parameters in Figure \ref{fig:cmdlfschem} and create similar narratives, and each one would highlight the importance of simultaneously fitting all uncertain quantities to obtain stellar model constraints.

In this study, we focus on the strength of convective overshooting of the stellar core ($\lambdac$), i.e., the distance in pressure scale heights (\hp) a convective element may pass beyond the convective zone. Core convective overshooting is an important and uncertain process that effects the central H fusion lifetimes of stars $\sim1.5-2.5\msun$, a fundamental quantity in stellar evolution.  Increasing the strength of core overshooting increases the main sequence luminosity for a given stellar mass, hence partially mimicking the effect of a younger cluster age in models with weaker core overshooting. 

This is the first in a series of papers from an HST archival program (AR-13901) to re-reduce and analyze $\sim150$ MC stellar clusters to obtain new constraints on stellar evolution models. Here we introduce a framework for using CMD-fitting to find the most probable stellar evolution model by simultaneously fitting 5 observational and model parameters while taking into account observational uncertainties and completeness. As a first exploration, we apply our framework to 6 LMC clusters with MSTO stars that are expected to have convective cores, therefore strong CMD signatures as a function of core overshooting strength. Our approach differs from typical isochrone fitting because we are able to use the posterior distribution functions (PDFs) to quantify the constraints on each parameter, as well as see any correlations between parameters, whether they are observationally or theoretically uncertain. 

In Section \ref{sec:convec} we discuss theory and existing measurements of core convective overshooting in stars. In Section \ref{sec:clusters} we discuss cluster selection, followed by Section \ref{sec:data} where we briefly describe the data acquisition, reduction, photometry, and artificial star tests. In Section \ref{sec:models} we describe the stellar evolution model grid we built, the CMD fitting software \match, our prior distributions, and results from CMD fitting using the model grid and mock data. In Section \ref{sec:analysis}, we discuss the derived cluster parameters when holding \lambdac\ at its canonical value and varying it. Finally, we conclude in Section \ref{sec:conclusions}. All magnitudes follow the VEGAMAG system.

\subsection{Previous Observational Constraints on Core Overshooting Strength}
\label{sec:convec}
The treatment of convection in the stellar interior affects the effective temperature, luminosity, and age of the MSTO in low-mass stars and the hot extension of the blue loop in intermediate-mass He-burning stars.  Core overshooting affects different parts of a CMD differently, depending on the age of the stellar population. Constraining core overshooting is important in astrophysics beyond the goal of precision stellar evolution models because uncertainties in core overshooting strength can be of order $5\%$ MS lifetime ($\tau_{\rm MS}$) for low mass stars. At MC metallicities, $5\%$ of a MS lifetime is a significant portion of subsequent evolutionary phases like the He-burning phase ($\sim20\%\ \tau_{\rm MS}$) and thermally pulsating AGB phase (TP-AGB; $\lesssim1\%\ \tau_{\rm MS}$). It is therefore a critical goal for those who study or use HB and TP-AGB models to push uncertainties in MS lifetimes smaller than the duration of the short-lived evolutionary phases after the MSTO.

Convection in stars is a complex, 3-dimensional time dependent process, and while efforts are underway to apply 3- and 2D models of convection to 1D stellar models \citep[e.g.,][]{Arnett2015}, these techniques are still too computationally expensive to be applied across all stellar ages and masses needed to synthesize stellar populations. Instead, convective energy transport in 1D stellar models typically follows the mixing length formalism \citep[MLT,][]{BohmVitense1958}, which defines the mixing length parameter $\alpha_{MLT}$, as the mean distance a convective element travels before being reabsorbed into its surrounding medium. The Sun is the main target for calibrating $\alpha_{MLT}$ \citep[e.g.,][]{Basu2009}.

The formalism to describe convective overshooting differs between stellar modeling groups. The main two varieties continue the formalism of MLT and parameterize the strength of convective overshooting in units of pressure scale height (\hp). The PARSEC models \citep{Bressan2012, Bressan2013} define the parameter \lambdac\ across the Schwarzschild boundary \citep{Bressan1981}, while other groups adopt the parameter \aov\ measured from above the Schwarzschild boundary. Whichever the preference, they can be compared following the relation, $\aov \sim \frac{\lambdac}{2} \hp$.

Observational constraints on core overshooting for masses $M\lesssim 3\msun$ historically come from by-eye fitting of isochrones to MSTO morphology of open clusters, after determining or adopting values for distance, reddening, and membership. The range of overshooting parameters fill in the range $0 \lesssim \lambdac\ \lesssim 0.5$, but are most commonly found to be 0.4 \citep[e.g.,][]{Demarque1994, Kozhurina-Platais1997, Sarajedini1999, Woo2001}. 
\citet{Bressan1993} suggested $\lambdac$ was not one value for all masses, and used $\lambdac=0.25$ for $1-1.5\Msun$ and $\lambdac$=0.5 for masses, M$\ge1.5\Msun$. A more gradual increase of the overshooting efficiency with mass was introduced by \citet{Demarque2004}.

In the early 2000's, ``by-eye'' isochrone fitting was gradually supplemented with more robust analyses based on the comparison with synthetic CMDs and luminosity functions \citep[e.g.][]{Woo2003, Bertelli2003}. This change in methods was enabled by the better photometric quality in the MCs with large telescopes and later with HST. These improvements were particularly important for studying MC clusters, which are in general more populated than their Galactic counterparts, and are often projected over sparely populated Galactic fields, hence reducing uncertainties related to low stellar counts and unknown membership probabilities. 

\citet{Woo2003}, using Yale isochrones \citep{Yi2001}, found the overshooting strength of $\lambdac \sim0.4\ \hp$ to best-fit the CMDs of the intermediate-age LMC clusters NGC~2173, SL~556, and NGC~2155 (cluster ages $\sim1.5-3$ Gyr; which would correspond to MSTO stars $\sim1.4-1.8$ \msun\ according to PARSEC models). Similar results were obtained also for NGC~2173 \citep{Mucciarelli2007}. \citet{Girardi2009} was able to simultaneously fit the dual red clump and MSTO in the center of the SMC cluster NGC~419 by adopting $\lambdac=0.47^{+0.14}_{-0.04}$ and $\log$ Age=$1.35^{+0.11}_{-0.04}$ Gyr (corresponding to a MSTO mass $\sim1.8$ \msun) and assuming uncertainties dominated by random errors. At higher masses, the young LMC cluster NGC~1866 has produced independent evidence of moderate core overshooting \citep{Barmina2002}, although the findings have been challenged \citep[see e.g.][]{Brocato2003}.

Eclipsing binaries have also been used to measure core overshooting, through core overshooting's effect on stellar radius \citep[e.g.,][]{Schroeder1997,Ribas2000}. The range in overshooting strength was found to be $0.48 \lesssim \lambdac\ \lesssim 0.64$ and increasing with increasing mass between $2.5-6.5\msun$. \citet{Claret2016} found contradictory results from reanalyzing a well measured set of 33 double-lined eclipsing binaries in the Milky Way, LMC, and SMC. They found \lambdac\ is independent of metallicity but depends on mass, such that \lambdac\ rises approximately linearly from 0 to 0.4 over the interval $1.2 \msun \le M \le 2.0 \msun$ and remains roughly constant for higher masses with a dispersion of $\sim0.06$ (their sample reaches 4.4 \msun).

With the burst of asteroseismology observations, new avenues for observational constraints have found some of the most extreme non-zero values of core overshooting, \citep[][$\lambdac=0.34\pm0.06 \hp$ for a Solar metallicity star with mass $M=0.95\msun$]{Deheuvels2010} and the highest \citep[][$\lambdac=2-2.5$ for Procyon, $M\sim1.5\msun$]{Guenther2014}.

In summary, studies that are focused on individual stars are converging on convective core overshooting increasing with increasing mass at least up to masses of 6.5 \msun, while constraints from stellar populations find \lambdac values between 0.4-0.5\hp. Reported uncertainties or dispersions of core overshooting strength are typically around the $15-20\%$ level or 0.06 \hp.

\section{Cluster Selection}
\label{sec:clusters}
\begin{figure}
\includegraphics[width=\columnwidth]{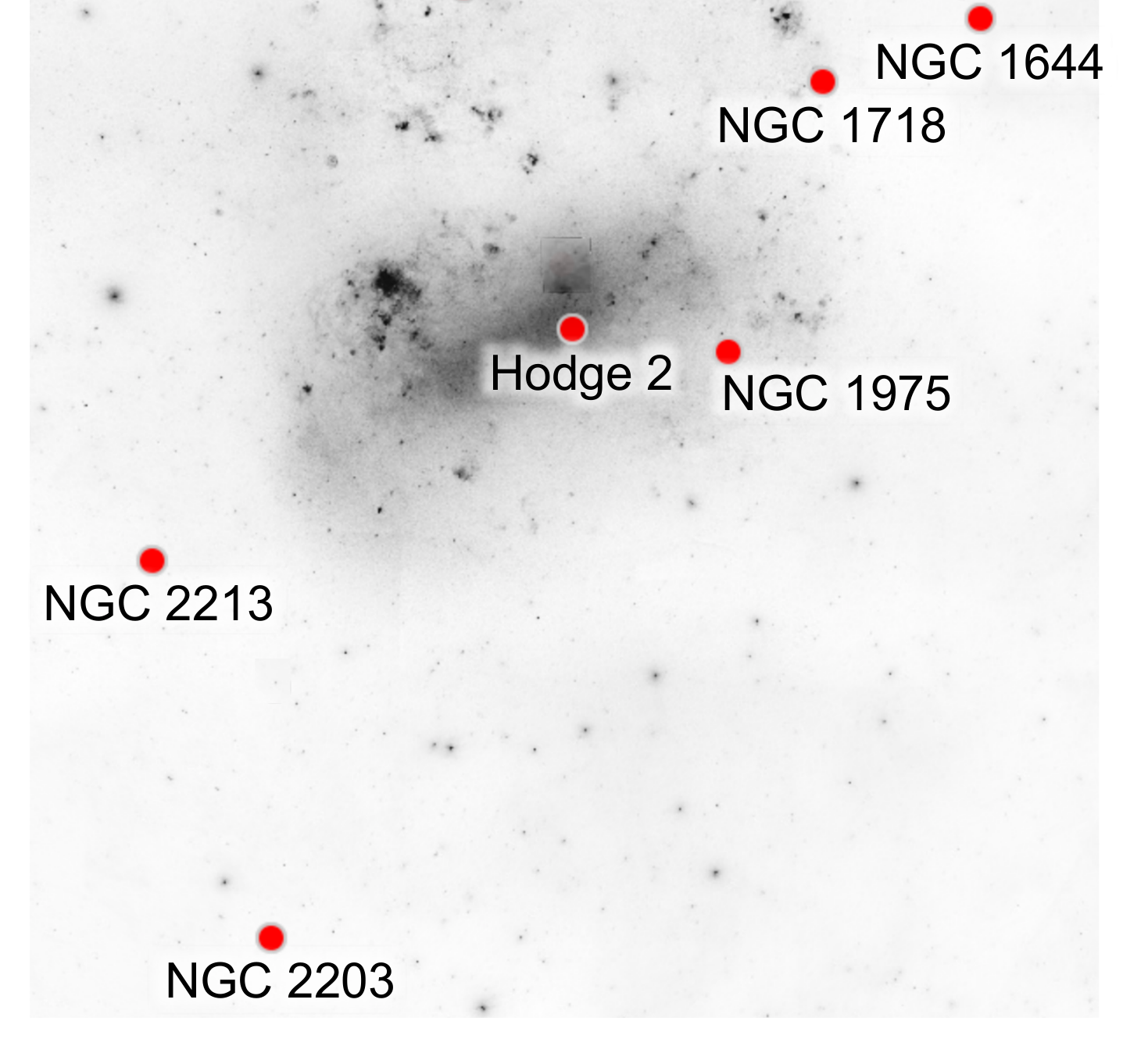}
\caption{Sky distribution of stellar clusters.}
\label{fig:radec}
\end{figure}

Clusters in our main program are described in detail in Fouesneau et al., (in prep). Briefly, we selected $\sim150$ clusters by cross matching MC cluster catalogs \citep{Bica2008, Baumgardt2013, Glatt2010} with the HST photometric archive (i.e., observations taken with ACS, WFC3, and WFPC2 in at least two optical wide-band filters). From this sample, we selected 6 clusters (HODGE~2, NGC~1718, NGC~2208, NGC~2213, NGC~1644, NGC~1795) with literature ages near 1.5 Gyr \citep{Bica2008} that were distributed throughout the LMC (see Figure \ref{fig:radec}). Cluster ages were chosen to be near 1.5 Gyr because their MSTO stars will have convective cores, and therefore, the strength of core overshooting will most dramatically affect the CMD morphology of the MSTO and red clump (discussed further in Section \ref{sec:effectcov} below). In effect, clusters were re-reduced from two HST programs: GO-9891 (PI: Gilmore) and GO-12257 (PI: Girardi).

The selected clusters have masses ranging from $\sim2.3\times10^4-1.3\times10^5\msun$ \citep{Baumgardt2013} and varying morphologies of the MSTO. Several clusters in the MCs have been discovered to have MSTOs that are extend in color and luminosity \citep[eMSTO; e.g.,][]{Milone2009} rather than MSTOs that show a narrow morphology that are typical in Galactic globular clusters \citep[c.f.,][]{Kalirai2012}.  Four of the six clusters in our sample have been identified as having eMSTOs, whose origins are actively being debated in the literature \citep[e.g.,][]{Milone2009, Goudfrooij2011, Goudfrooij2014, Correnti2014, Bastian2016}. For simplicity, we quote measured eMSTO widths as age spreads. Literature values of cluster properties are summarized in Table \ref{tab:lit}.

\begin{deluxetable*}{lllllllll}
\colnumbers
\tablecaption{Cluster Parameters from Isochrone Fitting in the Literature}
\tablehead{
\colhead{Name} &
\colhead{log Mass} &
\colhead{Metallicity} &
\colhead{$\mu_0$} &
\colhead{$A_V$} &
\colhead{Age} &
\colhead{eMSTO} &
\colhead{Ref.}  &
\colhead{Stellar}
\\
\colhead{} &
\colhead{\msun} &
\colhead{$Z$} &
\colhead{} &
\colhead{} &
\colhead{(Gyr)} &
\colhead{(Myr)} &
\colhead{} &
\colhead{Model}}
\startdata
Hodge 2 &  4.98 & $\bf{0.008}$      & 18.45          & 0.19          & 1.45          &     & 1 & PARSEC \\
        &       & $0.008$           & $18.40\pm0.03$ & $0.15\pm0.02$ & $1.30\pm0.05$ & 363 & 2 & Padua08 \\
\hline
NGC~1718 &  5.10 & $\bf{0.008}$      & 18.54          & 0.53          & 1.75          &     & 1 & PARSEC \\
        &       & $0.008$           & $18.42\pm0.03$ & $0.58\pm0.03$ & $1.80\pm0.05$ & 406 & 2 & Padua08 \\
        &       & $0.008^{+0.002}_{-0.001}$ & $18.73\pm0.07$ & $0.31\pm0.09$\tablenotemark{a} & $2.04\pm0.14$ & & 3  & Padua02 \\
\hline
NGC~2203 &  5.05 & $\bf{0.008}$      & 18.41          & 0.19                  & 1.75          &     & 1 & Padua08 \\
        &       & $0.008$           & $18.37\pm0.03$ & $0.16\pm0.02$         & $1.55\pm0.05$ & 475 & 2 & PARSEC \\
        &       & $\bf{0.006}$      & $\bf{18.49\pm0.09}$    & 0.34\tablenotemark{a} & $2.00\pm1.1$  &     & 4 & PARSEC\\
\hline
NGC~2213 &  4.56 & $\bf{0.008}$      & 18.40          & 0.16                  & 1.75          &     & 1 & PARSEC \\
        &       & $0.008$           & $18.36\pm0.03$ & $0.14\pm0.02$         & $1.70\pm0.05$ & 329 & 2 & Padua08 \\
        &       & $0.004\pm0.001$ & $18.56\pm0.08$ & $0.19\pm0.09$\tablenotemark{a} & $1.70\pm0.14$ & & 3 & Padua02 \\
        &       & $\bf{0.006}$      & $\bf{18.49\pm0.09}$     & 0.34\tablenotemark{a} & $1.78\pm1.1$ & & 4 & PARSEC \\
\hline
NGC~1644 &  4.32 & $0.008$           & 18.48          & 0.03\tablenotemark{a} & 1.55          & $<50$ & 5 & BaSTI\\
\hline   
NGC~1795 &  4.36 & $0.008$           & 18.45          & 0.31\tablenotemark{a} & 1.3           & $<50$ & 5 & BaSTI \\
\enddata
\tablenotetext{a}{$E(B-V)$ values were converted to $A_V$ assuming $R_V=3.1$}
\tablecomments{Reference for Column 2: \citet{Baumgardt2013}. References and fitting notes for Columns 4--7: 1) \citet{Niederhofer2016} -- $\mu_0$ and $A_V$ ``by eye"; 2) \citet{Goudfrooij2014}; 3) \citet{Kerber2007}; 4) \citet{Piatti2014}; 5) \citet{Milone2009} -- ``by hand.'' Stellar model references in Column 9: Padua08 -- \citet{Marigo2008}; Padua02 -- \citet{Girardi2002}; PARSEC -- \citet{Bressan2012}; BaSTI -- An August 2008 version of \citet{Pietrinferni2004}. Bold values denote fixed quantities during isochrone fitting. Uncertainties included when available.}
\label{tab:lit}
\end{deluxetable*}

\section{Archival Observations}\label{sec:data}
\subsection{Photometry and Reduction}
\begin{figure*}
\begin{center}
\includegraphics[width=0.3\textwidth]{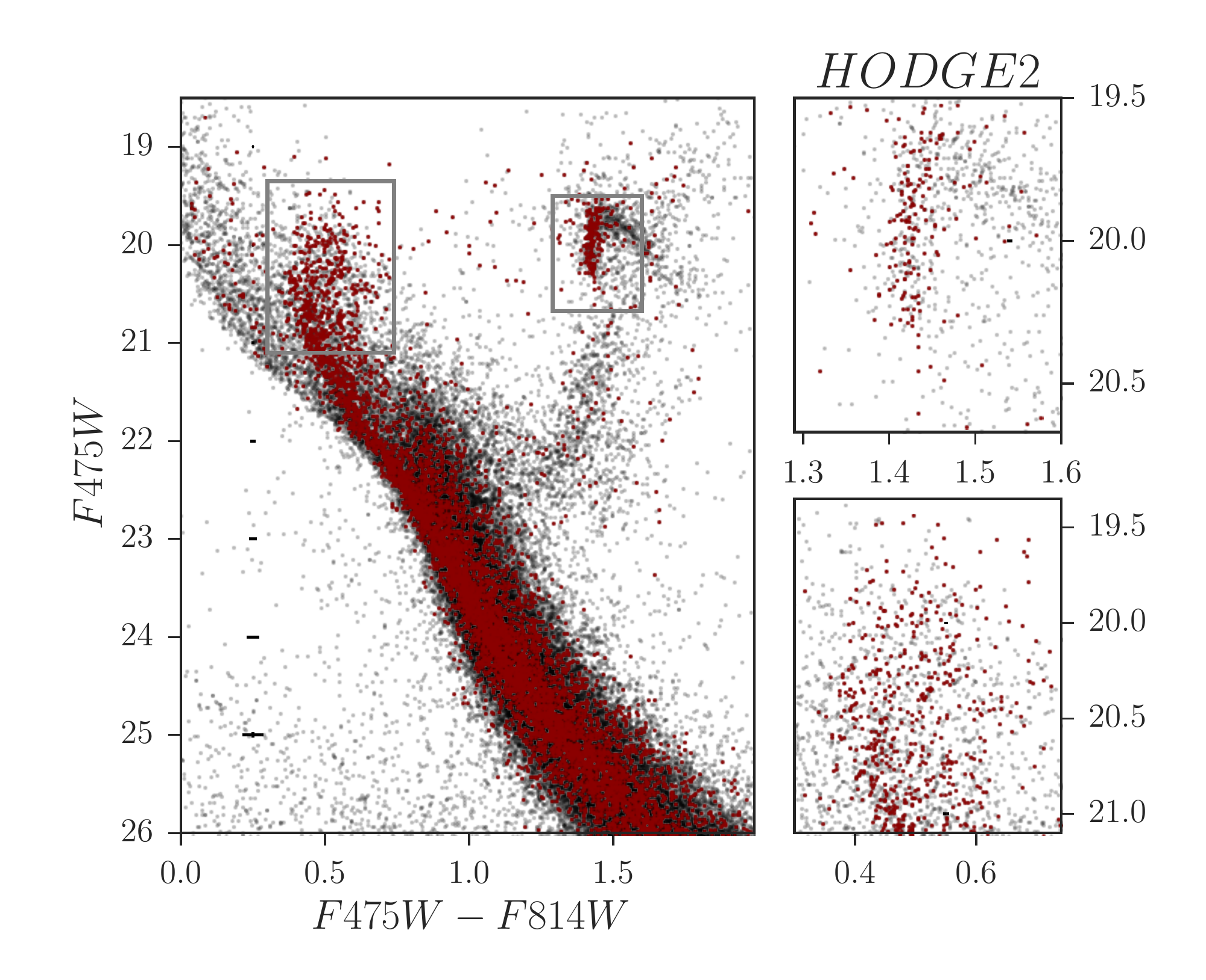}
\includegraphics[width=0.3\textwidth]{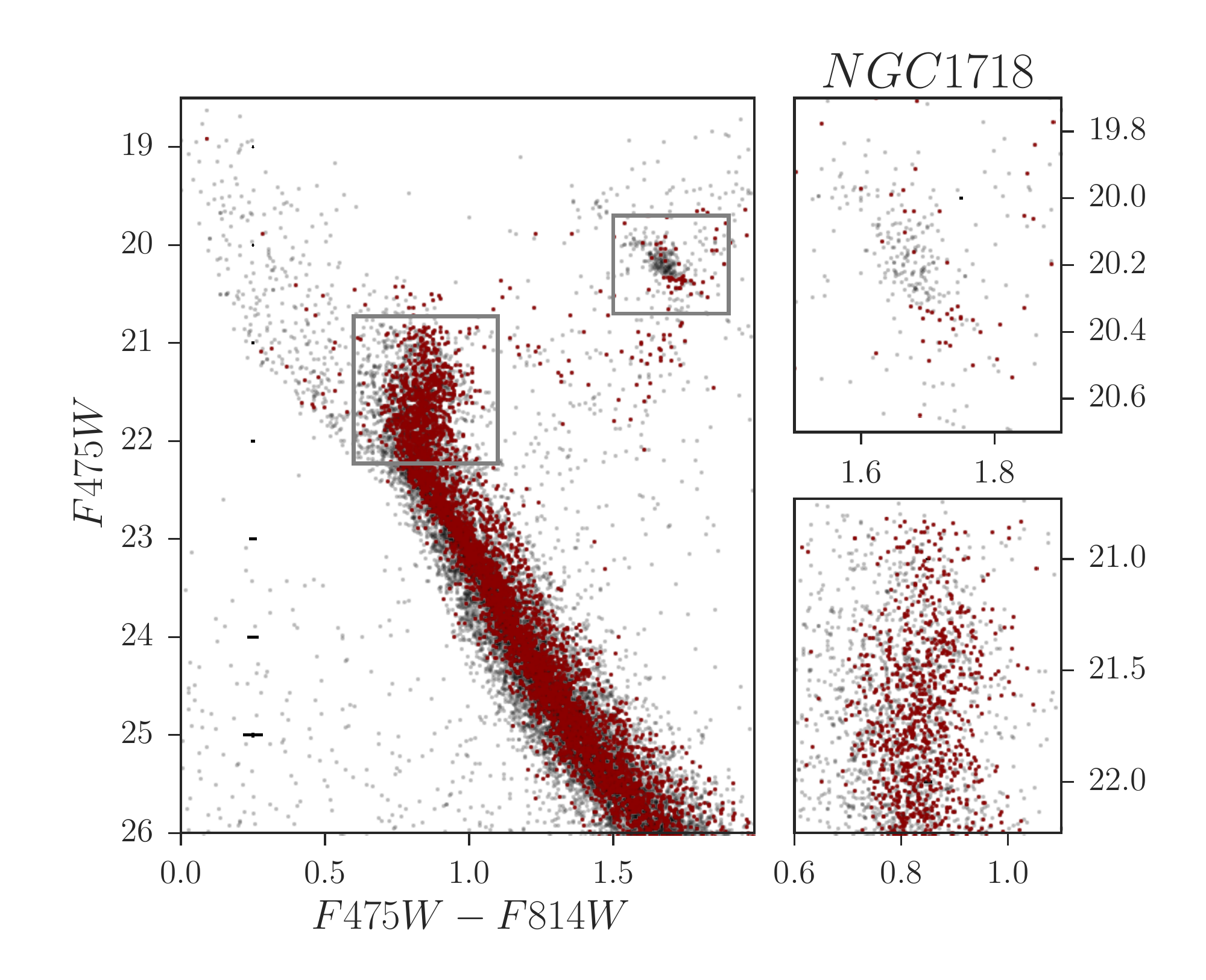}
\includegraphics[width=0.3\textwidth]{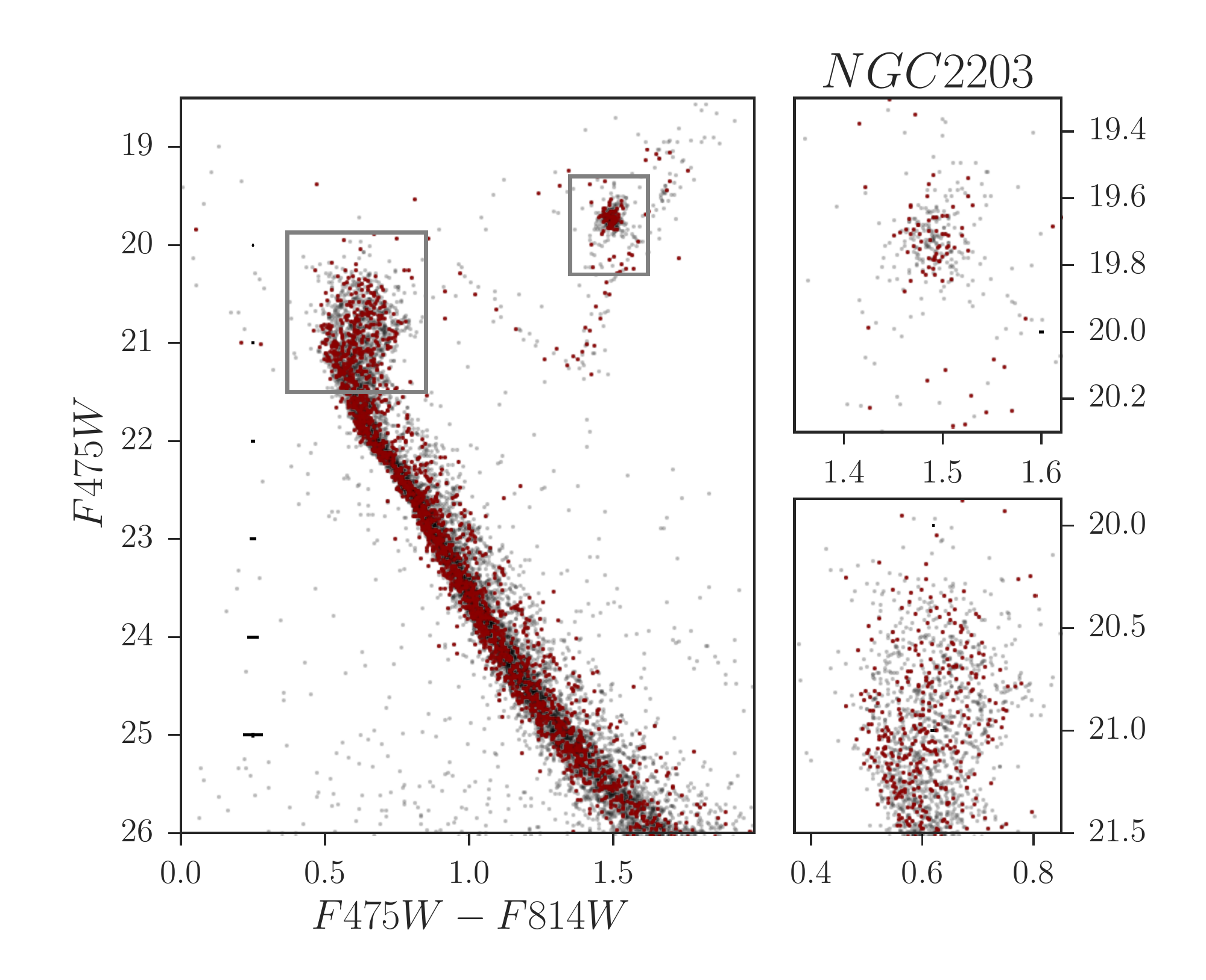}
\includegraphics[width=0.3\textwidth]{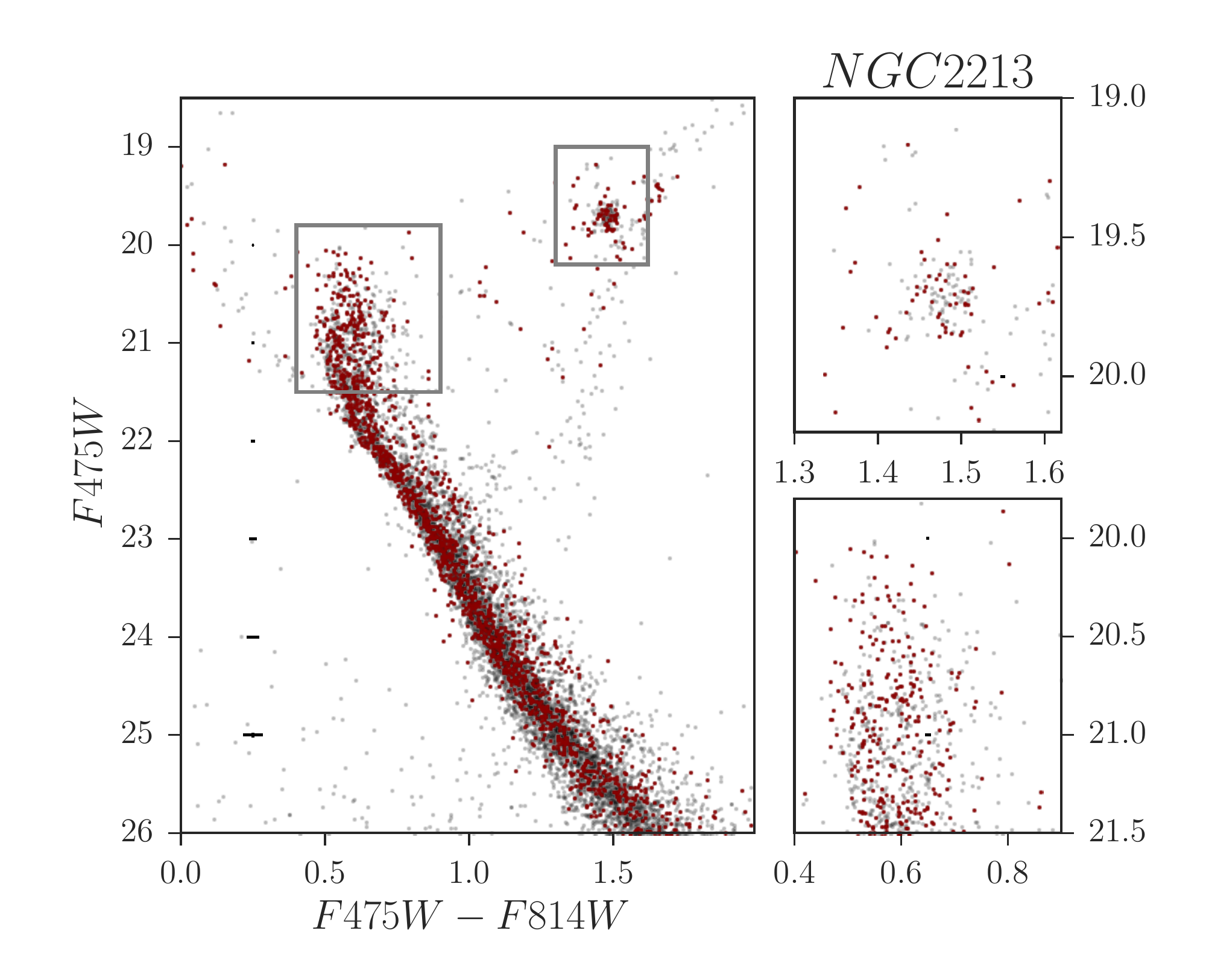}
\includegraphics[width=0.3\textwidth]{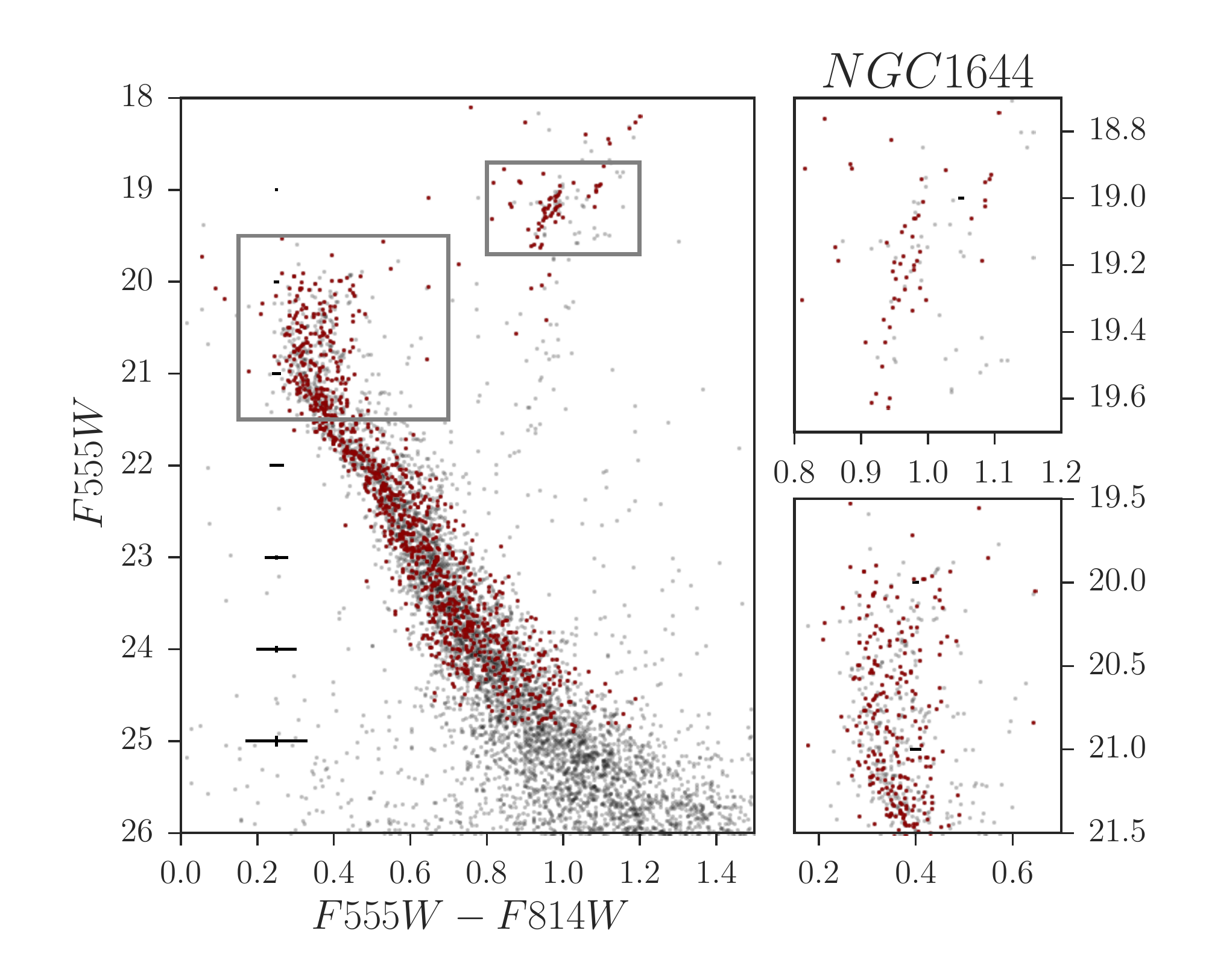}
\includegraphics[width=0.3\textwidth]{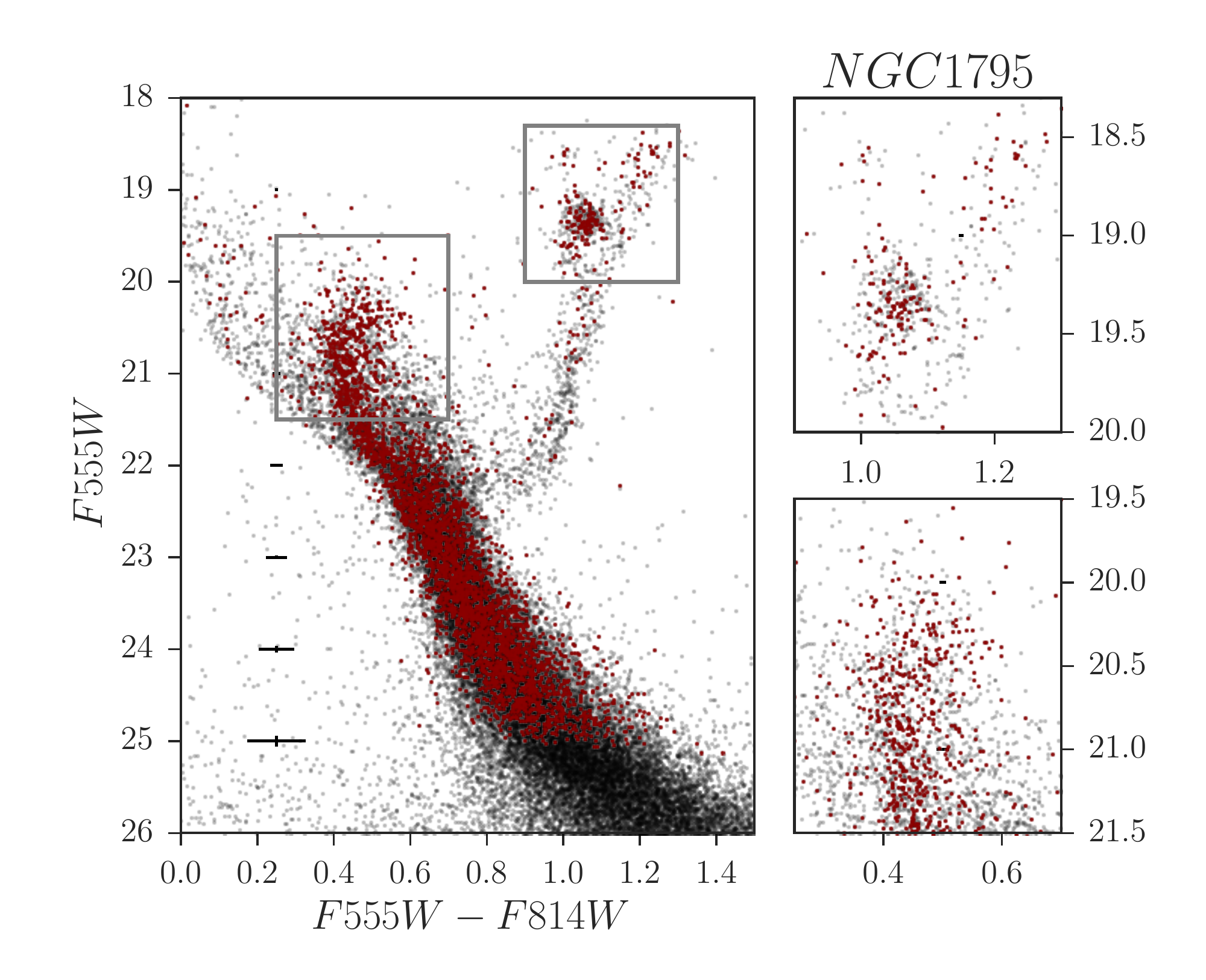}
\caption{CMDs with insets MSTO (bottom) and HB (top) for each cluster. Black points are the full field {\tt gst} catalog (Section \ref{sec:data}) and red points are the stars with at least 70\% cluster membership probability (Section \ref{sec:priors}). Mean photometric uncertainties are shown on the left side of each CMD.}
\end{center}
\label{fig:cmds}
\end{figure*}

ACS and WFC3 archival data were re-reduced using the University of Washington data reduction pipeline which was developed to reduce large HST programs, e.g., ANGST and PHAT \citep{Dalcanton2009, Dalcanton2012}. Its current capabilities are described in detail in \citet{Williams2014}. Briefly, {\tt flt} and {\tt flc} images were downloaded from the Mikulski Archive for Space Telescopes (MAST; flt for WFC3/IR; flc for ACS and WFC3/UVIS), astrometrically aligned using the cross-camera alignment software developed for PHAT as part of {\it astrometry.net}, and cleaned of cosmic rays using the {\it{astrodrizzle}} package \citep{Gonzaga2012}. We then used DOLPHOT \citep{Dolphin2000} for PSF photometry and created three photometric catalogs, {\tt phot}, {\tt st}, and {\tt gst}, and we use the {\tt gst} catalogs for our analysis. These catalogs provide 3 different levels of measurement quality.  The {\tt phot} catalogs are the closest to the full photometric output table from DOLPHOT. The {\tt st} catalogs are culled from the {\tt phot} catalogs and are limited to $S/N \ge 4$ and {\tt sharpness}$^2$ values in at least one filter to be below 0.2, 0.15 for ACS and UVIS respectively. The {\tt gst} catalogs are a subset of the {\tt st} catalogs that have {\tt crowding} values below 1.3. CMDs of the {\tt gst} catalogs are shown in black in Figure \ref{fig:cmds} and the archival data are summarized in Table \ref{tab:archival}. Full details of our data reduction pipeline are postponed to our instrument paper (Fouesneau et al., in prep).

\begin{deluxetable*}{lp{2cm}lllp{2cm}l}
\tablecaption{HST Archival Data}
\tablecolumns{6}
\tablewidth{0pt}
\tablehead{
\colhead{Target} &
\colhead{Proposal ID} &
\colhead{Instrument} &
\colhead{Filter} &
\colhead{Exposure Time (s)} &
\colhead{Date}
}
\startdata
HODGE~2	& 12257 & WFC3/UVIS & F814W & 1430 & Jan 21, 2012 \\
HODGE~2	& 12257 & WFC3/UVIS & F475W & 1440 & Jan 21, 2012 \\
NGC~1718	& 12257 & WFC3/UVIS & F814W & 1430 & Dec 02, 2011 \\
NGC~1718	& 12257 & WFC3/UVIS & F475W & 1440 & Dec 02, 2011 \\
NGC~2203	& 12257 & WFC3/UVIS & F814W & 1980 &  Oct 08, 2011 \\
NGC~2203	& 12257 & WFC3/UVIS & F475W & 1520 &  Oct 08, 2011 \\
NGC~2213	& 12257 & WFC3/UVIS & F814W & 1430 &  Nov 29, 2011 \\
NGC~2213	& 12257 & WFC3/UVIS & F475W & 1440 &  Nov 29, 2011 \\
NGC~1644	& 9891	& ACS/WFC & F555W & 250    & Oct 07, 2003 \\
NGC~1644	& 9891	& ACS/WFC & F814W & 170    & Oct 07, 2003 \\
NGC~1795	& 9891	& ACS/WFC & F555W & 200    & Aug 09, 2003 \\
NGC~1795	& 9891	& ACS/WFC & F814W & 300    & Aug 09, 2003 \\
\enddata
\tablecomments{Uniformly reduced archival observations retrieved from MAST.}
\label{tab:archival}
\end{deluxetable*}

\subsection{Artificial Star Tests}
In order to characterize the photometric errors and completeness of the HST data, we placed $\sim$100k artificial stars for each cluster.  Artificial stars are distributed rather uniformly in CMD space covering the full magnitude and color range of the data, and weighted such that fainter mags have relatively larger numbers of tests. Artificial stars are distributed spatially according to a King profile, literature values for center and half-light radius, fixed concentration, covering a range in radius out to four half-light radii, and bounded by ACS or UVIS field of view. 

\subsection{Cluster Parameters}
We fed our {\tt gst} catalogs into \asteca\ \citep{asteca}, an automated stellar cluster analysis package, to estimate the cluster parameters and cluster membership. Table \ref{tab:asteca} lists the derived cluster centers, radii, and great circle distances between derived cluster centers and values from \citet{Bica2008}.

\begin{deluxetable*}{lllll}
\tablecaption{\ASteCA-derived cluster parameters}
\tablehead{
\colhead{Cluster} &
\colhead{$\alpha$ Center} &
\colhead{$\delta$ Center} &
\colhead{$r_{\rm{cluster}}$} &
\colhead{$\Delta_{\rm{center}}$\tablenotemark{a}} \\
\colhead{} &
\colhead{(J2000)} &
\colhead{(J2000)} &
\colhead{arcsec} &
\colhead{arcsec}
}
\startdata
HODGE~2  & 5 17 48.816$\pm$0.048 & -69 38 41.640$\pm$0.720 & $33.84\pm 1.80$ & 1.24 \\
NGC~1718 & 4 52 25.704$\pm$0.072 & -67 03 05.040$\pm$1.080 & $38.88\pm 2.52$ & 4.06 \\
NGC~2203 & 6 04 43.392$\pm$0.072 & -75 26 19.320$\pm$1.080 & $42.12\pm 2.16$ & 6.17 \\
NGC~2213 & 6 10 42.240$\pm$0.072 & -71 31 45.840$\pm$1.080 & $29.52\pm 2.16$ & 2.13 \\
NGC~1644 & 4 37 39.792$\pm$0.096 & -66 11 55.680$\pm$1.440 & $26.28\pm 2.52$ & 5.25 \\
NGC~1795 & 4 59 47.280$\pm$0.096 & -69 48 03.960$\pm$1.440 & $57.60\pm 2.88$ & 6.58 \\
\enddata
\tablenotetext{a}{Separation between \citet{Bica2008} center coordinates and \ASteCA-derived cluster center coordinates.}
\label{tab:asteca}
\end{deluxetable*}

Full details of the fitting algorithms are in the main \asteca\ paper, in short, the cluster centers are determined by the maximum spatial density using a two-dimensional Gaussian kernel density estimator. The cluster radius is set to where the radial density profile becomes indistinguishable from the background stellar density. Contamination from non-cluster stars within the cluster radius are discussed in Section \ref{sec:cmdfitting}.  Stars within the radius of the cluster and have at least 70\% membership probability are used as input photometry and shown in red in Figure \ref{fig:cmds}.

\section{Methods}\label{sec:models}
\subsection{Stellar Evolution Models}
The Padova-Trieste Stellar Evolution Code \citep[PARSEC][]{Bressan2012, Bressan2013} is a major update to the Padua models \citep{Girardi2000}. PARSEC adopts the solar metallicity value of $\zsun=0.01524$ and the scaled solar distribution of elements heavier than $^4$He are taken from \citet{Grevesse1998} except for Li, C, N, O, P, S, K, Fe, Eu, Hf, Os, and Th, which are taken from \citet{Caffau2011} and references therein. The initial He abundance ($Y_i$) for each metallicity set is calculated based on the primordial He abundance, $Y_p$=0.2485, \citep{Komatsu2011} and the Helium-to-metals enrichment ratio, $\Delta Y/ \Delta Z$=1.78, which was obtained in \citet{Bressan2012} using solar values. That is, $Y_i = Y_p + (\Delta Y / \Delta Z)\ Z_i$. \citep[see Section 2 and Table 1 of][]{Bressan2012}.

PARSEC adopts an overshooting prescription that linearly increases in strength from no overshooting to the a maximum value ($\Lambda_{\rm{max}}$) between two mass steps ($M_{O1}, M_{O2}$). The Padua models \citep{Girardi2000,Bertelli2008} set $\Lambda_{\rm{max}}=0.5$, $M_{O1}$ = 1.0\msun, and $M_{O2} = 1.5\msun$ at all metallicities. In PARSEC V1.2S, $\Lambda_{\rm{max}}$ is the same, however, the mass steps are derived separately for each metallicity and Helium content.  

\subsubsection{The PARSEC Core Overshooting Model Grid}
We relax the PARSEC setting of $\Lambda_{\rm{max}}=0.5$ and calculate a grid of 3,560 stellar evolution tracks using PARSEC V1.2S \citep[updates from][]{Chen2014, Tang2014} beginning at the pre-main sequence and ending at the termination of He-burning. Core overshooting in our model grid is not only calculated for Hydrogen-burning (MS) cores, but Helium-burning cores (HB or HeB) are also calculated with the labeled core overshooting strength. Table \ref{tab:stelmods} lists the details of the stellar model grid.

\begin{deluxetable}{lc}
\tablecaption{PARSEC Convective Core Overshooting Model Grid}
\tablehead{
\colhead{Parameter} &
\colhead{Values}}
\startdata
Mixture ($Z_i, Y_i$) & 0.0005, 0.249 \\
& 0.0010, 0.250 \\
& 0.0020, 0.252 \\ 
& 0.0040, 0.256 \\
& 0.0060, 0.259 \\ 
& 0.0080, 0.263 \\
& 0.0100, 0.267 \\ 
$\Lambda_c$ (\hp) & $0.3-0.6$: $\Delta \lambdac = 0.1$ \\
Mass ($\msun$) & $ 0.1 \le M \le 2.4:  \Delta M  \le 0.05\ \msun$ \\
 & $ 2.6 \le M \le 6.4:  \Delta M  = 0.20\ \msun$ \\
 & $ 7.0 \le M < 12.0:  \Delta M  = 1.0\ \msun $ \\
 & $ 12.0  < M \le 20.0:  \Delta M  = 2.0\ \msun$ \\
\enddata
\label{tab:stelmods}
\tablecomments{We interpolated the overshooting grid to obtain \lambdac=0.45, and 0.55 models and extended the grid to \lambdac=0.80 for  select clusters.}
\end{deluxetable}

\subsubsection{The Effect of Core Overshooting}
\label{sec:effectcov}
Increasing core overshooting strength allows more fresh nuclear material to fuse, making for a larger and hotter core that leads to a longer main sequence or HB lifetime and a brighter and cooler MSTO. We now describe how such a physical change in the interior of a star is expected to affect observations of a nearly single age stellar population. First, we illustrate how the core fusion lifetimes change, followed by the Hertzsprung-Russell diagram (HRD) and CMD appearance of sample stellar evolution tracks chosen at masses relevant to the ages of the sample clusters, and finally three simple synthetic stellar populations as snapshots of different epochs in star formation. As we will show, longer core fusing lifetimes translate to higher stellar densities on the CMD than otherwise expected. A brighter and cooler MSTO changes the expected position of a star on a CMD, leading to possible misidentification of a star's mass, age, or distance.

\begin{figure}
\includegraphics[width=\columnwidth]{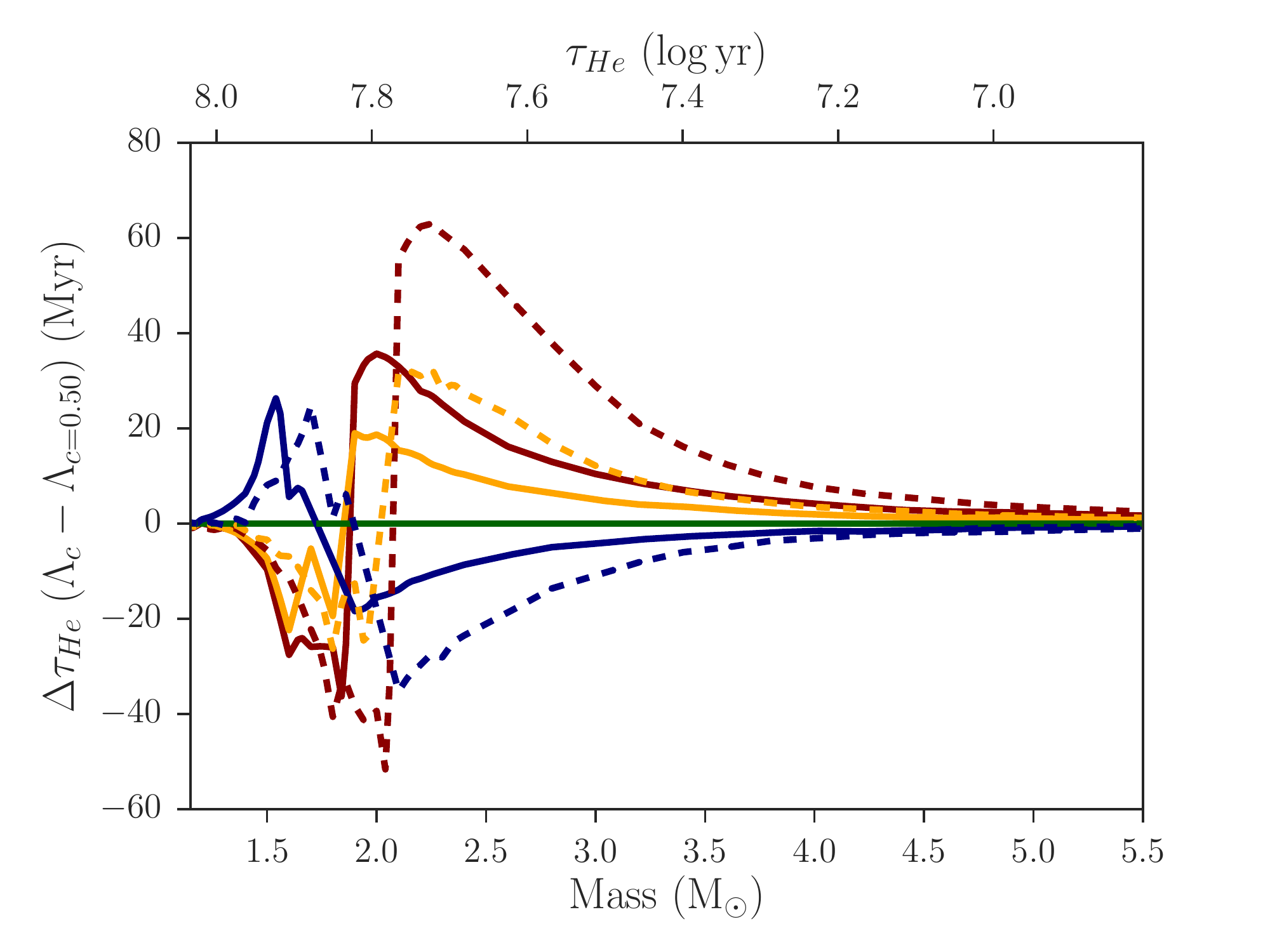}
\includegraphics[width=\columnwidth]{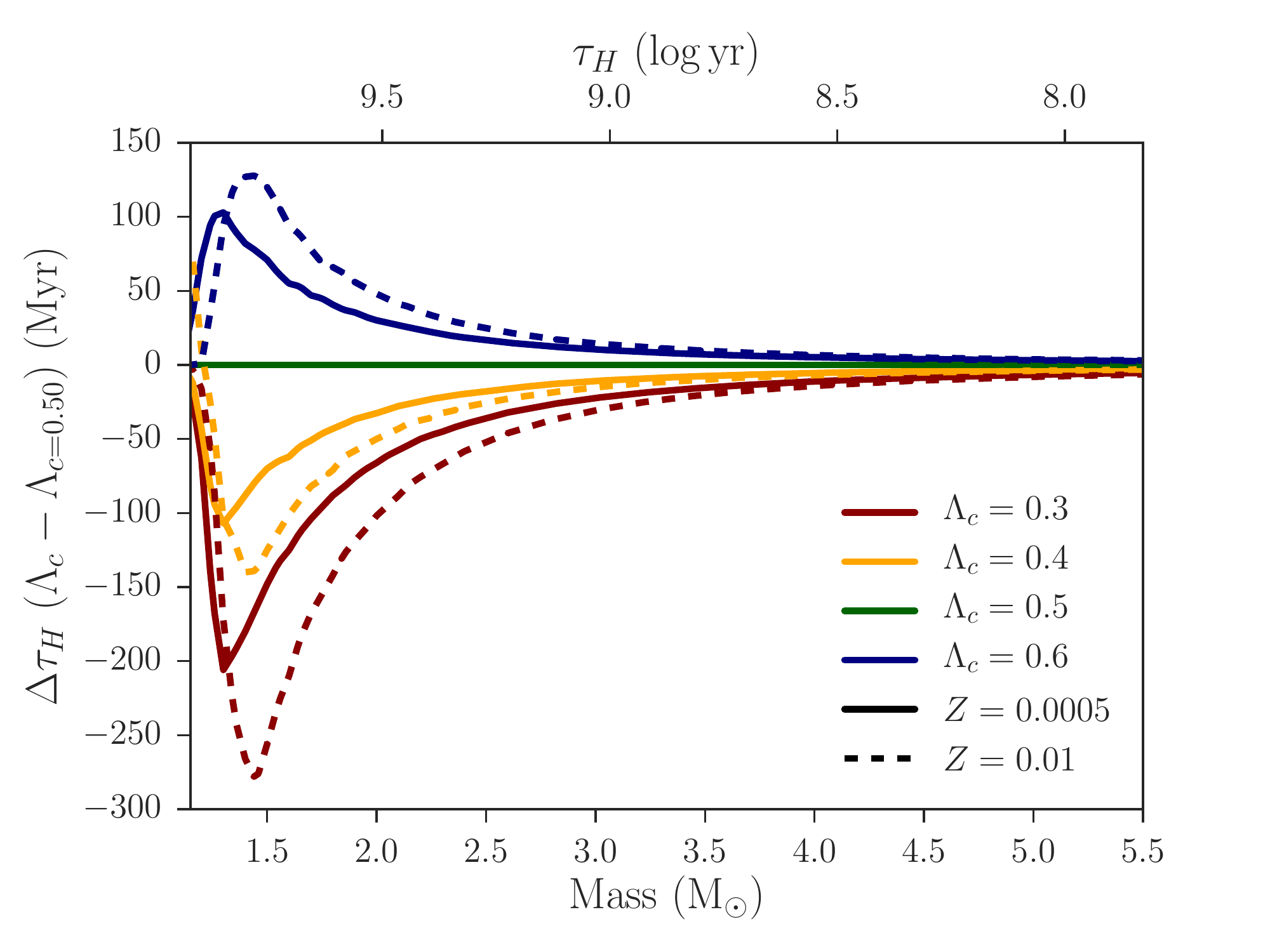}
\caption{Increasing core overshooting increases Helium burning (top) and main sequence (bottom) lifetimes at masses around 1.5 \Msun, when the convective core is largest. Shown are the differences, as a function of mass, in Hydrogen burning lifetime (i.e, MSTO age) and Helium burning lifetime of varying levels core overshooting compared to the canonical PARSEC value of $\lambdac=0.50\ \hp$. The solid and dashed lines denote the lowest and highest metallicities in the core overshooting model grid. Shown as a guide on the top axes of each panel are the H or He burning lifetimes for Z=0.008, $\lambdac=0.50\ \hp$ models. Differences in MSTO lifetimes quickly increase until they peak at $\lambda_{\rm max}$, and increase once again for Helium burning stars that begin fusing Helium in a non-degenerate state. A comparison to other stellar modeling groups is discussed in Appendix \ref{appx:othermodels}}
\label{fig:COV_MSTO}
\end{figure}

Figure \ref{fig:COV_MSTO} shows as a function of mass, the difference in core fusing lifetimes of core overshooting strength compared to the canonical PARSEC value ($\lambdac=0.5\ \hp$). A solid and dashed line mark the extrema of the stellar evolution model grid metallicities and are plotted for each calculated overshooting value (a similar comparison to other stellar modeling groups is discussed in Appendix \ref{appx:othermodels}). The effect of core overshooting on MSTO age quickly increases from from low masses to a peak, which is set by the linear ramp-up of core overshooting from the lowest masses in PARSEC. Soon after $\Lambda_{\rm{max}}$ is reached, the effect of the convective core on MSTO and He burning age decreases, as expected, with increasing mass as core convection becomes less important. As an example, for a 1.5 \msun\ star, an increase in core overshooting of $\Delta \lambdac\ = 0.1\ \hp$ leads to a $\sim 100-150$ Myr ($\sim5\%$) longer MS lifetime depending on metallicity. A $\sim 100-150$ Myr increase in MS lifetime is a small effect for the MS, however, for a 1.5 \msun\ star, it is longer than the entire core He burning lifetime, and longer than the thermally pulsating AGB lifetime \citep[e.g.,][]{Rosenfield2016}.

\begin{figure*}
\includegraphics[width=0.5\textwidth]{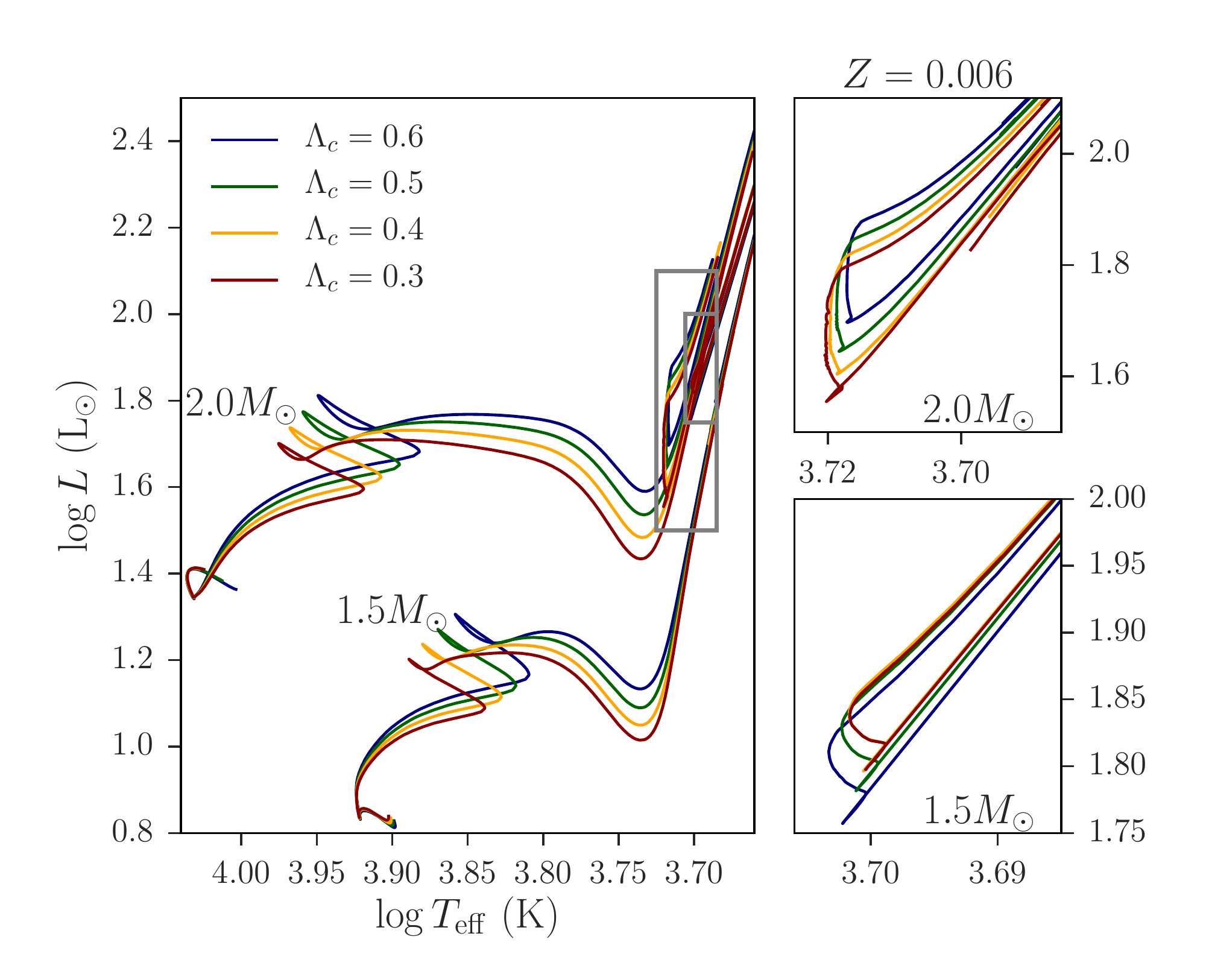}
\includegraphics[width=0.5\textwidth]{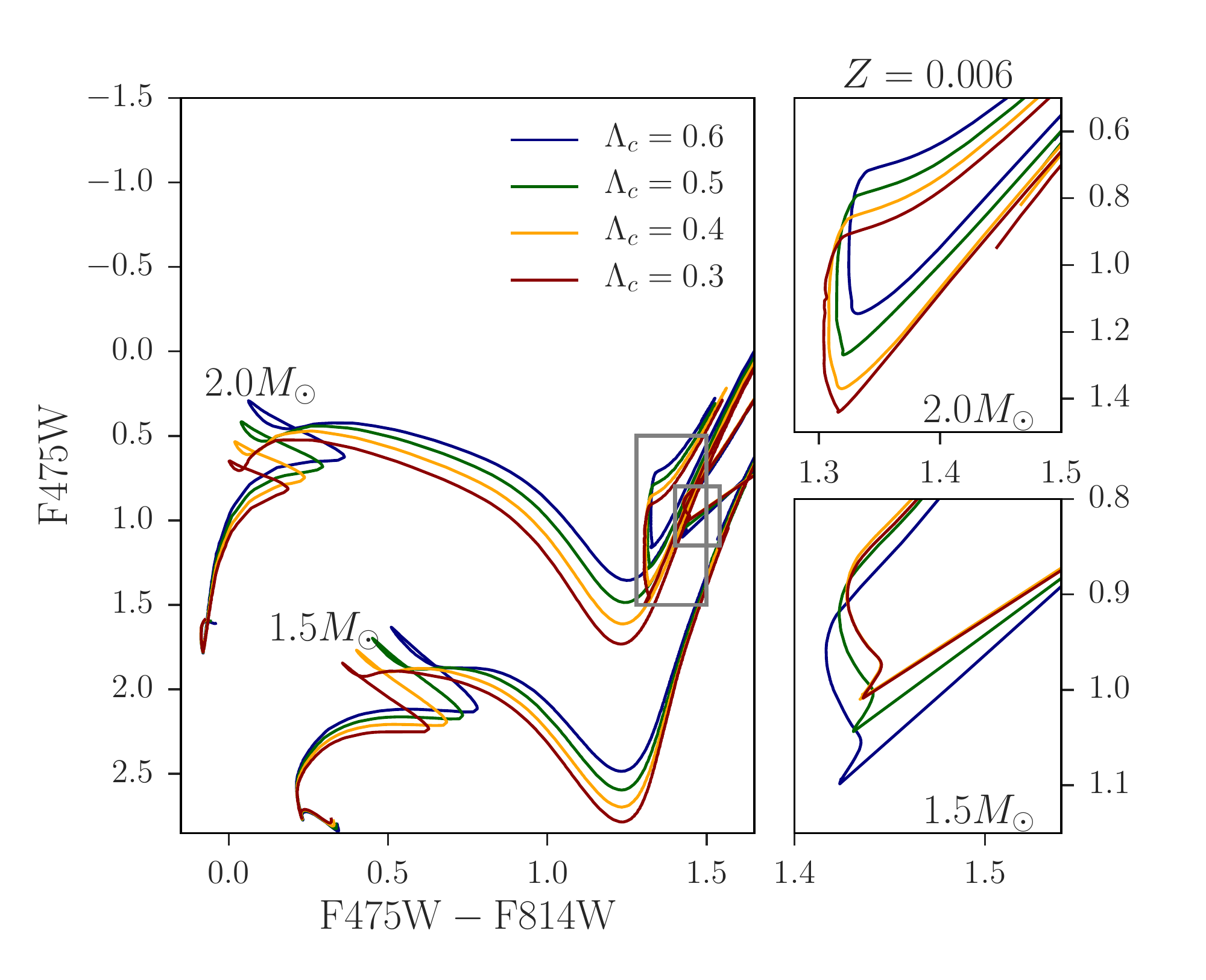}
\caption{Left: Hertzsprung-Russell diagram (left) and CMD (right) showing model stellar evolutionary tracks at two masses with varying levels of core overshooting strength. Increasing core overshooting increases the luminosity and effective temperature of the MSTO to a lesser and lesser degree with increasing mass, changes the morphology of the SGB, and decreases the extent of the red clump. Insets: expanded views showing the Helium burning phases for each mass.}
\label{fig:COV_HRD}
\end{figure*}

Figure \ref{fig:COV_HRD} shows an HRD (left panel) and a CMD (right panel) of evolutionary tracks from the core overshooting grid selected at initial masses of 1.5\msun\ and 2\msun. The transformation from the HRD to magnitudes and colors are based on the tables of bolometric corrections from \citet{Girardi2008} (revised for the latest ACS/WFC3 filter transmission curves) and rely on the ATLAS9 atmospheric models from \citet{Castelli2004}. The same transformations are also implemented in the \MATCH\ routines (see Section \ref{sec:models}).

The MSTO is shifted to brighter and to cooler effective temperatures with increasing core overshooting. However, the amplitude of the brighter and cooler excursions decreases with increasing mass. There are also clear morphological differences around the MSTO, subgiant branch, and helium burning phases. The extension between the minimum effective temperature on the MS and the MSTO increases with increasing core overshooting, the luminosity dip after the MSTO is more pronounced with increasing core overshooting, and the extent to hotter temperatures of the Helium burning tracks decreases with increasing core overshooting. 

\begin{figure*}
\includegraphics[width=.25\textwidth]{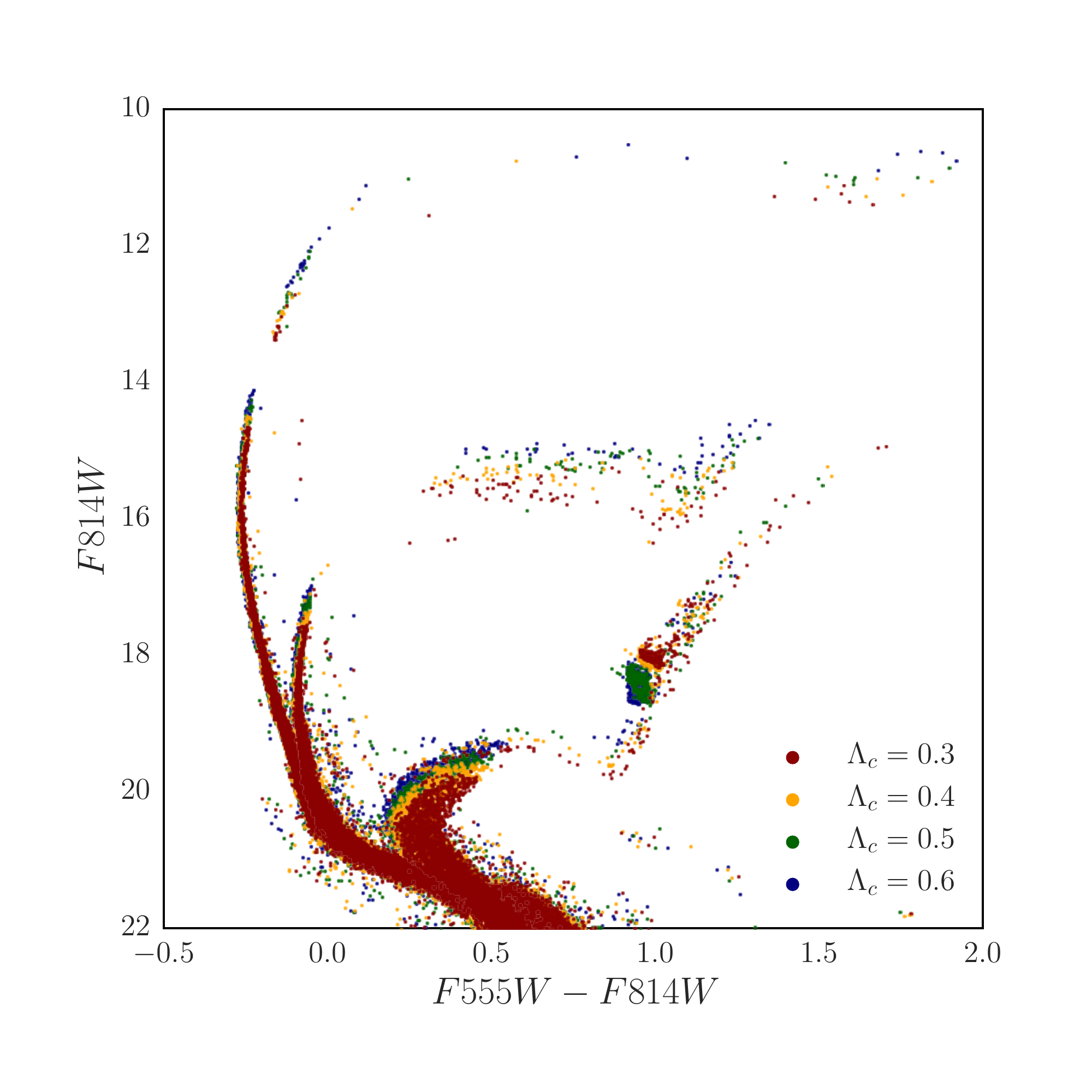}
\includegraphics[width=.24\textwidth]{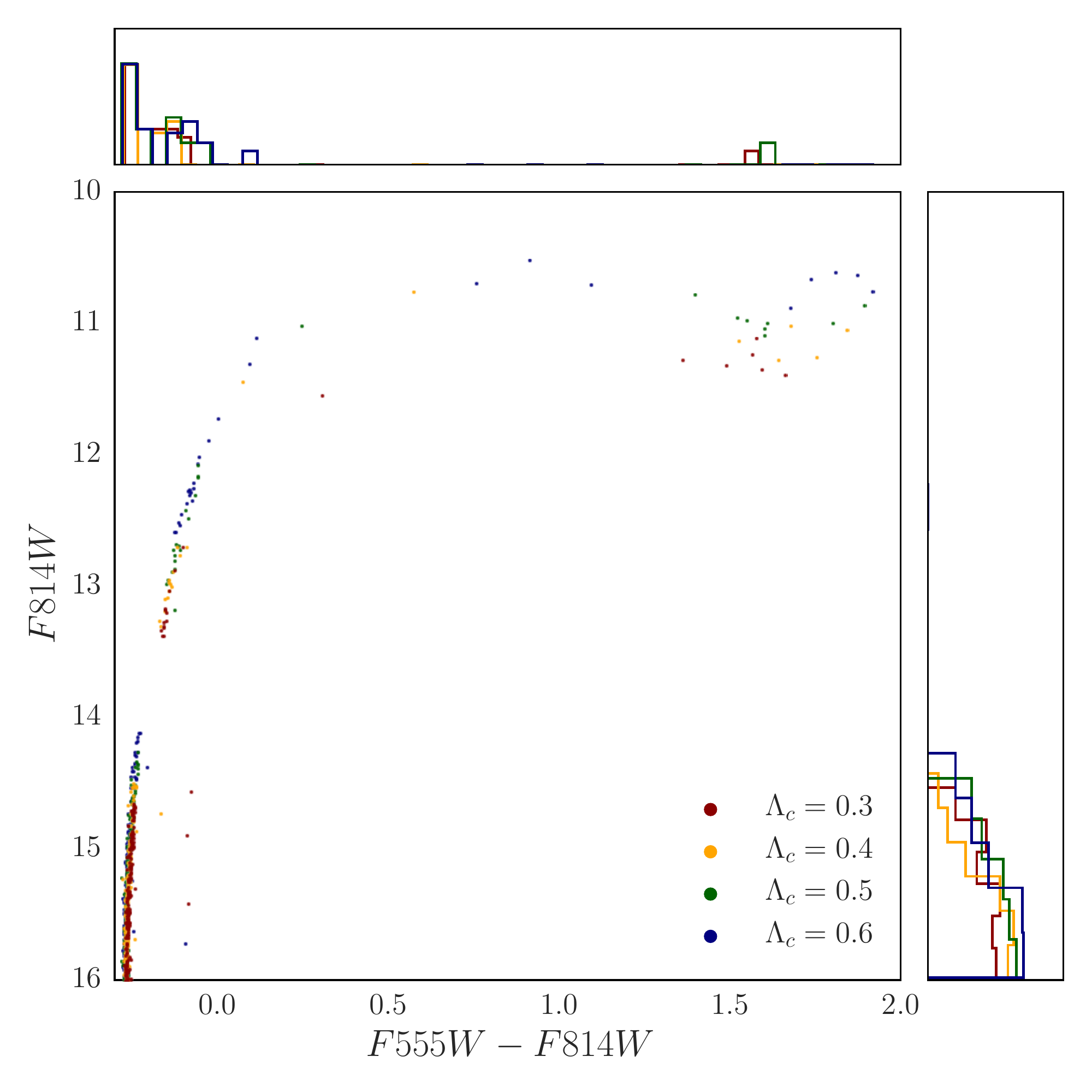}
\includegraphics[width=.24\textwidth]{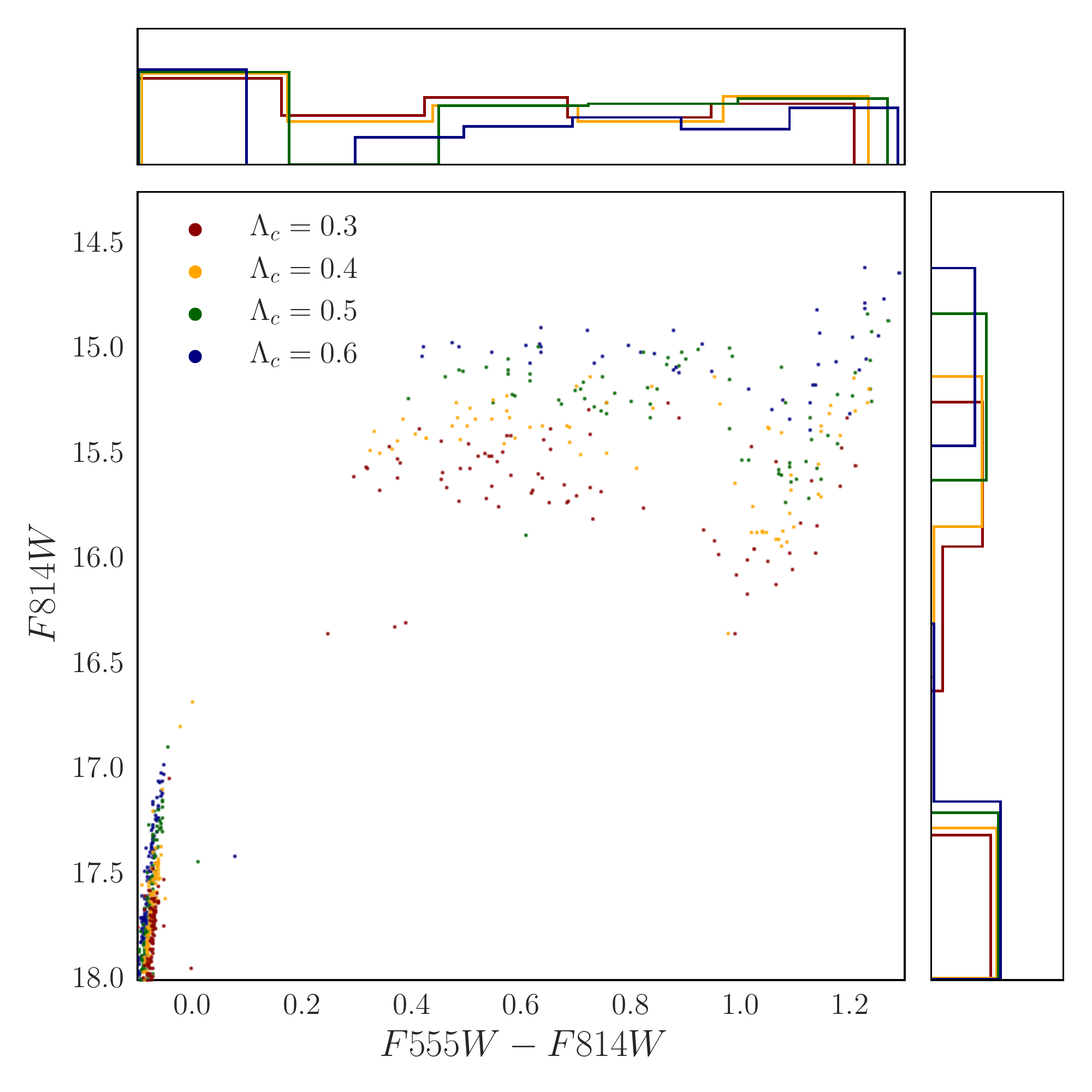}
\includegraphics[width=.24\textwidth]{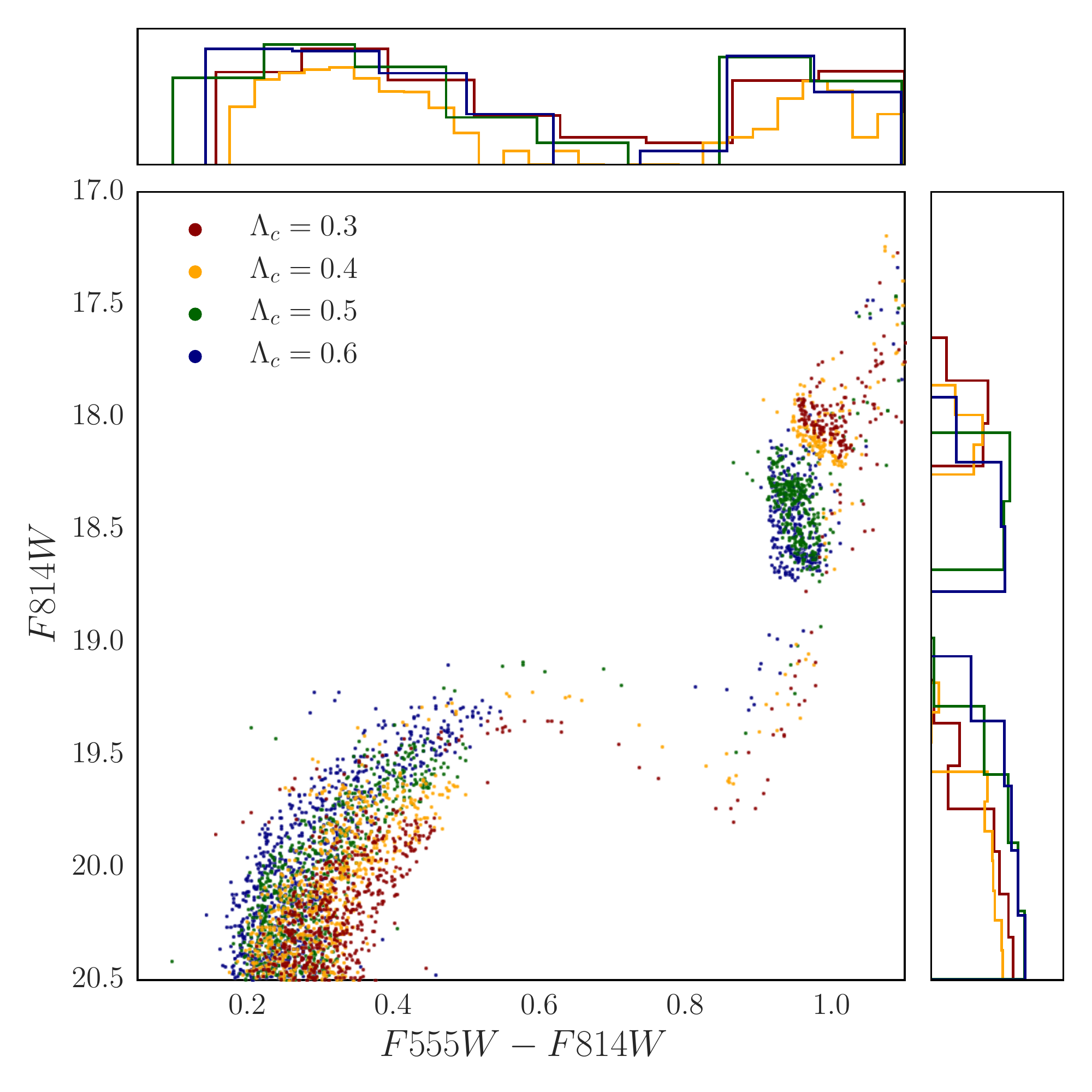}
\caption{CMDs of synthetic stellar populations calculated at a distance of $\mu_0=18.5$, starting at three ages (14 Myr, 180 Myr, and 1.4 Gyr from left to right) and SF lasting 60 Myr, four values of core overshooting strength (red, yellow, green, purple for $\lambdac= 0.3, 0.4, 0.5$, and 0.6), and no binaries or extinction. Left-most panel is a combined summary of the right panels. Top and side axes of each of the three right panels show color and magnitude histograms. Each CMD shows increasing core overshooting strength increases the brightness of the MSTO. Other differences in morphology due to core overshooting differ with population age. In the center two panels, younger populations have fewer but brighter blue and red core He burning stars (clumps of stars brighter than than the MS and bluer or redder than F555W-F814W$\sim0.75$) with increasing overshooting. In the right panel, the morphology of both the MSTO and the RC differ with increasing overshooting strength as the MSTO and SGB are brighter while the RC is fainter and more populated.}
\label{fig:fake_cmds}
\end{figure*}

The age differences and the morphological changes in the stellar evolution tracks due to core overshooting strength culminate in the shape and number density of a stellar population on a CMD. Figure \ref{fig:fake_cmds} shows synthetic stellar populations from models of each core overshooting strength produced using the {\tt fake} routine in the \MATCH\ package (see Section \ref{sec:cmdfitting}). The synthetic stellar populations are made of one burst of constant star formation lasting 60 Myr starting at 14 Myr (center left panel), 180 Myr (center right panel), and 1.4 Gyr (right panel), a constant initial metallicity of $Z=0.006$ (corresponding to $[M/H]=-0.40$~dex), a \citet{Salpeter1955} IMF, a distance $\mu_0=18.50$, typical photometric uncertainties, and neglecting binaries and extinction.

Figure \ref{fig:fake_cmds} shows that observational signatures of core overshooting differ at different epochs of star formation. In the younger populations (center two panels) the core helium burning stars show the most significant differences. Increasing overshooting strength decreases the extent of the blue loop, which manifests in the youngest CMD as brighter red and blue helium burning stars (RHeB, BHeB; these stars have masses $>12\msun$ and $>3\msun$ in the center-left and center-right panel, respectively) with increasing overshooting. The numbers of RHeB and BHeB also decreases with increasing overshooting strength, which is more significant at the youngest aged population than the intermediate aged population.

For the oldest aged synthetic stellar population (right panel), exhibits different MSTO, sub-giant branch (SGB), and red clump (RC) morphologies. Decreasing core overshooting strength increases the MSTO color, increases the number of SGB stars, lowers the total number of RC stars, and leads to a brighter, more compact RC.

\subsection{CMD Fitting}
\label{sec:cmdfitting}
CMDs are powerful tools for understanding the history of star formation in stellar populations. A CMD can be well approximated by a linear combination of bursts of star formation over cosmic time \citep[][]{Dolphin2002}. 
Exploiting the tenet, the \match\ software package \citep[][and refs. therein]{Dolphin2016}, specifically, the {\tt calcsfh} module was designed to derive the most likely SFH from a binned CMD (Hess diagram) of the photometry of a mixed-age stellar population.

To compare an observed Hess diagram to a model Hess diagram, \match\ first constructs the model Hess diagram given an input set of stellar models and user-specified prior on the IMF slope, binary fraction, metallicity, metallicity dispersion, and color and magnitude bin sizes. {\tt calcsfh} will then iterate over distance, extinction, and epoch of SF burst, until either the most probable linear combination of ages is found (for mixed-age stellar populations), or until the likelihood is calculated for each epoch of SF (for near-single age stellar populations). We iterate calls to {\tt calcsfh} to such that the single value priors become a distribution. We now describe how we set these prior distributions.

\subsubsection{Prior Distributions}
\label{sec:priors}
Table \ref{tab:search} lists our model priors and CMD fitting grid resolution. We chose our priors to be uninformative and flat distributions over a range set beyond derived literature values but limited for computational efficiency. We constrained values (IMF slope, binary fraction) that only effect the lower MS, where the photometric uncertainties are highest.

\begin{deluxetable*}{lcc}
\tablecaption{Priors and the {\tt calcsfh} grid search space}
\tablehead{
\colhead{Parameter} &
\colhead{Range} &
\colhead{Step Size}}
\startdata
IMF ($\Gamma$) & 1.35 & \citep[fixed]{Salpeter1955} \\
Binary Fraction & $0.30$ & \citep[See][fixed]{Milone2009, Milone2016} \\
Distance ($\mu_0$; mag) & $18.3-18.7$\tablenotemark{a} & 0.05 \\
Extinction ($A_V$; mag) & $0.0-0.5$\tablenotemark{a} & 0.05 \\
Age (Gyr) & $1.0-2.5$ & 0.06 \\
Metallicity (\feh; dex) & $-0.85 - -0.15$ & 0.10 \\
Core overshooting strength ($\lambdac$; \hp) & $0.3-0.6$\tablenotemark{b} & 0.1; 0.05 between \lambdac=0.4-0.6 \\
Color (mag) & $\sim0.0-2.0$ (varies by cluster)  & 0.05 \\
Magnitude (mag) & $\sim16-24$ (varies by cluster) & 0.10 \\
\enddata
\tablenotetext{a}{NGC~1718 distance modulus prior was 18.5-18.9, and its extinction prior was 0.0-1 with the same listed step sizes.}
\tablenotetext{b}{We extended the core overshooting grid to \lambdac=0.80 for NGC~1718 and NGC~2203}
\label{tab:search}
\end{deluxetable*}

\paragraph{Binary Fraction}
Observations of pre-MS stellar systems \citep[e.g.,][]{Kroupa2011}, N-body simulations \citep[e.g.,][]{Marks2011a} and theoretical arguments \citep[e.g.,][]{Goodwin2005} suggest most stars are likely formed in binaries. The cluster radius, stellar type, cluster density, and age, among other factors contribute to the binary fraction. \citet{Sollima2007} found the binary fraction varies from 0.1-0.5 for low-density Galactic globular clusters. Galactic field populations have measured binary fractions from $\sim0.2-0.8$ \citep[e.g.,][and refs. therein]{Marks2011}, with the fraction decreasing with decreasing stellar mass. In the MCs, \citet{Milone2009, Milone2016} determined the binary fraction ranges from $\sim0.19-0.46$ for several MC clusters. We set the binary fraction to the approximate median found in the \citeauthor{Milone2009} papers, 0.3, with a uniform mass ratio distribution from $0.1-1.0\msun$. We will explore variations of binary fractions in subsequent work that includes LMC and SMC clusters of differing age. 

\paragraph{IMF Slope}
We do not attempt to constrain the low mass MS stars in this study, and adopt the \citet{Salpeter1955} IMF slope of $\Gamma=1.35$. The lowest mass stars to be included in our analysis have M = 0.8 \msun. 

\paragraph{Distance}
We adopt a true distance modulus range of $\mu_0=18.30-18.70$~mag and step size 0.05 mag, which encompasses common literature values of $\mu_0=18.36-18.54$~mag with the exception of NGC~1718, which has a derived literature distance of $18.73\pm0.07$ \citep{Kerber2007}. Therefore, we extended the distance modulus range to 18.9~mag for NGC~1718 to ensure the best fitting distance was not at the edge of the grid.

\paragraph{Extinction}
Following the method to set our distance priors, values of $A_V$ from the literature range from  $0.03-0.58$~mag. With $A_V$ step size of 0.05 mag, we set our prior limits from $0-0.6$~mag, again extending the grid edge for NGC~1718 to 1.0~mag.

\paragraph{Age}
Clusters were selected because their literature ages were around 1.5 Gyr, we limited the age prior to $1.5\pm0.75$ Gyr for computational efficiency.

\paragraph{Metallicity}
Most isochrone fitting of the clusters in our sample set the metallicity to either Z=0.008 (\feh=$-$0.28~dex) or Z=0.006 (\feh=$-$0.4~dex). We set our prior limits to Z=$0.002-0.01$ (\feh=$-0.85- -0.15$~dex) with a step size of 0.1 dex.

\paragraph{CMD Range and Binning}
Using the simulated stellar populations described in Section \ref{sec:sens}, we ran {\tt calcsfh} setting the color and magnitude bin sizes at all combinations of values 0.01, 0.05, 0.1, and 0.15 mag. We confirmed the heuristic tenet from \citet{Dolphin2002}: CMD bin sizes should be set smaller than the important observed CMD features, in our case, the MSTO and the HB. We adopt the color bin size of 0.05 mag and magnitude bin size of 0.10 mag.

\paragraph{Cluster Contamination}
In the \asteca\ package, the user may calculate the star-by-star probability of cluster membership by invoking a non-parametric Bayesian decontamination algorithm (DA) based on the method of \citet{Cabrera1990} which was originally applied to open clusters. We limit the input photometry to stars within the cluster radius with at least 70\% membership probability (shown in red in Figure \ref{fig:cmds}). 

\paragraph{Age and Metallicity Resolution}
For stellar clusters, it is useful to measure the goodness of data-model fit of a simple stellar population (SSP) as a function of age. The minimum possible SSP age resolution in \match\ is set by an internal pre-compiled grid of partial CMDs, for our core overshooting grid this resolution is d\feh=0.05 dex and dlog Age=0.01 (log yr). This high resolution grid allows us to test SF in age bins $\gtrsim 20$ Myr at ages of 1 Gyr (2\%). 
However, we found there was no added improvement in the fitting between resolution of 2\% and 6\%  so we adopted an SSP age resolution of 60 Myr as it provided an optimal balance between computational time and sensitivity to cluster age spreads. 

\subsection{Resolving Core-Overshooting Strength in Synthetic Populations}
\label{sec:sens}
We have seen that varying core overshooting strength propagates to CMD in ways that depend on age, which is the manifestation of the importance of convection in the stellar core (see Section \ref{sec:effectcov}). On an optical CMD, a stellar population that is brighter and cooler could mean that it is in fact closer and has less extinction than assumed. For example, it is tempting to point out the $F814W$ mag of the MSTO of the youngest synthetic population in Figure \ref{fig:fake_cmds} shows $\sim0.3$ mag spread depending on the strength of core overshooting. These would correspond to bright stars with negligible photometric uncertainty. However, in these optical filter sets and at that population age, a large 0.3 spread in core overshooting strength at the MSTO is nearly indistinguishable from a $\sim$0.2 mag uncertainty in the distance modulus. This underscores the importance of using the entire CMD to test models of uncertain evolutionary phases. It also behooves us to test the sensitivity of CMD-fitting and our core overshooting grid.

We ran the stellar population synthesis module {\tt fake} within \match\ to simulate simple stellar populations. The {\tt fake} module takes as input the same user-specified parameters as listed in Section \ref{sec:cmdfitting}, including artificial star tests to convolve with the model Hess diagrams and returns a synthetic CMD. We ran {\tt fake} to simulate a constant burst of SF at 1.5 Gyr $\pm\ 30$ Myr, \feh=$-$0.40~dex, a metallicity dispersion of 0.10 dex, at each grid value of core overshooting strength, and convolved the CMD with a typical cluster artificial star test uncertainty profile.

Using the {\tt MATCH fake} photometric catalogs as input to {\tt calcsfh}, we derived the best fitting CMD by searching over interstellar extinction, age, metallicity, and core overshooting strength (see Table \ref{tab:search}). In all cases, that is, for each of the four mock data input catalogs, {\tt calcsfh} clearly recovered the input parameters.

\subsubsection{Likelihood}
The best fitting model is found by minimizing the Poisson-equivalent of $\chi^2$ \citep[see][]{Dolphin2002}:
\begin{equation}
\label{eqn:starprob}
\chi^2_P = 2 \sum m_i - n_i + n_i \ln \left(\frac{n_i}{m_i}\right)
\end{equation}
Where $m_i$ is the number of model points and $n_i$ is the number of data points in the Hess diagram bin $i$.

To visualize the likelihoods, we produce marginalized posterior distribution functions (PDFs) of each parameter, and  joint-marginalized PDFs for each parameter pair. Joint-marginalized PDFs are comparable $\chi^2$ maps for Gaussian distributions, in our case they are $\chi_P^2$ maps.

The posterior distributions provide the full story of the uncertainties and correlations between parameters, given the PARSEC models and our priors. However, it is useful to note the most probable value and estimate the uncertainties for each fitted parameter. To do so, we report the ``best fit'' as the maximum posterior probability and take the 16th and 84th quartiles of a polynomial fit to posterior distribution as uncertainties. For a Gaussian distribution, these values would correspond to the mean and $1\sigma$.

\subsection{Systematic Uncertainties Due to \lambdac} 
To explore the effect of core overshooting strength on the derived cluster parameters, we marginalized the mock data results over the true value of \lambdac\ (i.e., \lambdac\ of the input synthetic stellar population) and the assumed value of \lambdac\ (i.e., \lambdac\ used to derive the cluster parameters). 

\begin{figure*}
\includegraphics[width=\textwidth]{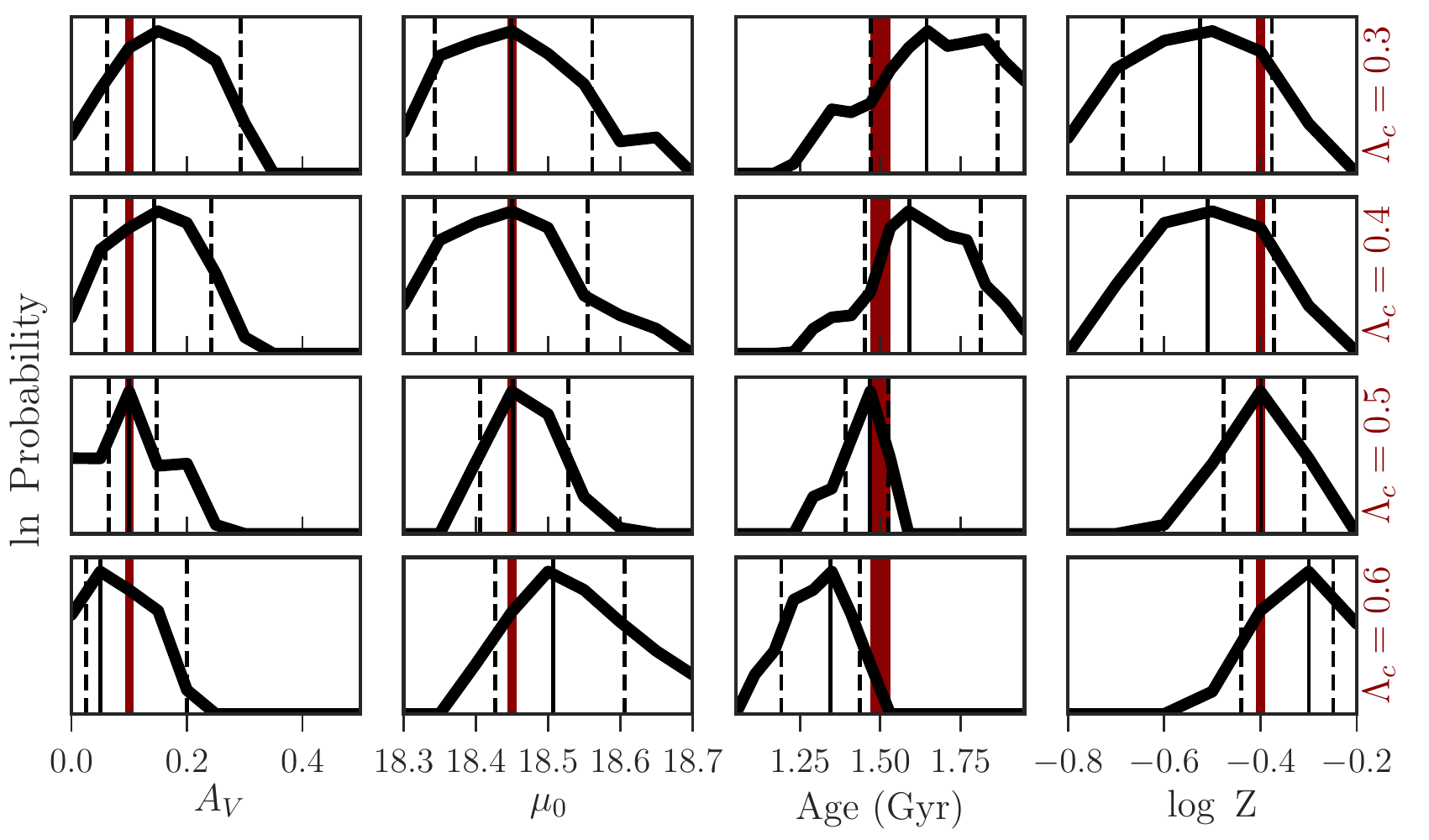}
\caption{Marginalized PDFs of synthetic stellar populations derived with the canonical value of \lambdac=0.50 \hp. The input stellar population parameters (truth) are shown in red, including the \lambdac\ value noted on the right-most panels. Dashed lines mark the 16th and 84th quartiles of the polynomial fit to the distribution and solid vertical black lines mark the maximum posterior probability. Incorrect assumptions on convective core overshooting strength will systematically offset derived cluster parameters. For a population with a mean age of 1.5 Gyr and SF lasting 60 Myr, uncertainties of order $\pm0.05\ \hp$ in \lambdac\ will introduce systematic offsets of $A_V\sim-0.04$ mag,  $\mu_0\sim+0.002$ mag,  Age $\sim-120$ Myr, and log Z $\sim+0.01$ dex, with increasing \lambdac\ (see text).} \label{fig:sspmpdf}
\end{figure*}

Figure \ref{fig:sspmpdf} shows the marginalized PDFs derived with an assumed value of \lambdac=0.50 and all calculated true values. Red vertical lines show the (true) input values used to create the synthetic stellar populations, including the \lambdac\ values noted on the right vertical axes.

The systematic offsets introduced as a function of increasing true \lambdac\ follow  from  the discussion in Section \ref{sec:effectcov} and Figures \ref{fig:COV_MSTO}-\ref{fig:fake_cmds}. For example, we have seen that increasing core overshooting strength increases core fusion lifetimes, therefore underestimating core overshooting strength will  bias  derived cluster ages older.  This effect can be seen comparing the age panels in top row and the third row of Figure \ref{fig:sspmpdf} (the third row being where the assumed \lambdac\ matches the true \lambdac). We can further estimate the systematic offsets as a function of core overshooting strength expected for a population aged $\sim$1.5 Gyr. The median offsets of the maximum posterior probabilities are $A_V\sim-0.04$ mag,  $\mu_0\sim+0.002$ mag,  Age$\sim-120$ Myr, and log Z$\sim+0.1$ dex, when increasing \lambdac\ by $0.1 \hp$.  In other words, for intermediate-aged stellar clusters, distance and extinction seem to be immune from uncertainties of core overshooting strength, however, age and metallicity may have disconcertingly large systematic offsets when \lambdac\ is uncertain by more than $\pm0.05 \hp$.

\begin{figure*}
\includegraphics[width=\textwidth]{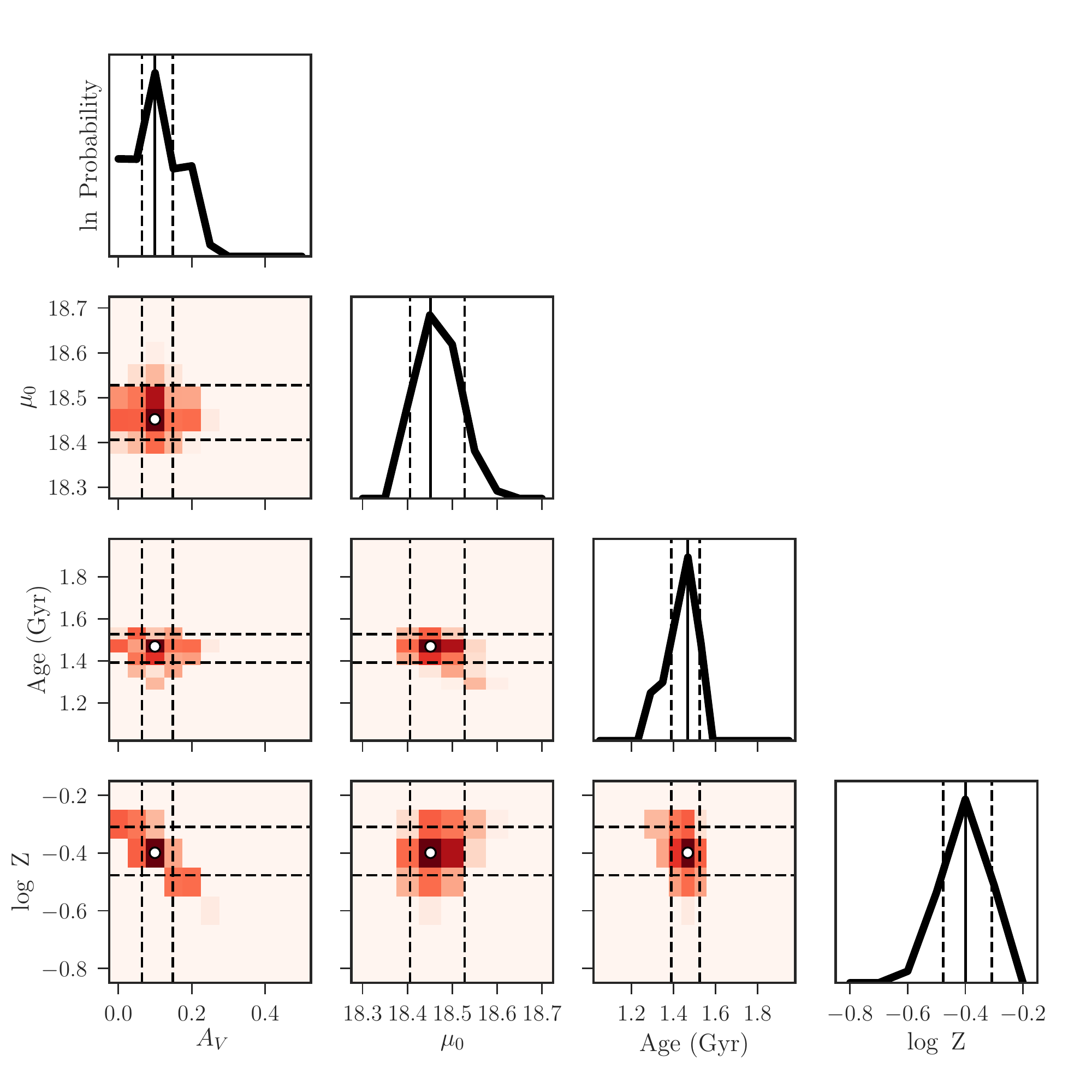}
\caption{Joint-marginalized PDFs of a 1.5 Gyr $\pm\ 30$ Myr stellar population calculated with \lambdac=0.5 \hp\ and fit assuming \lambdac=0.5\ \hp (see third row of Figure \ref{fig:sspmpdf}). See appendix \ref{appxovis5} for PDFs with true $\lambdac$ values mismatching the assumed value.}\label{fig:sspjmpdf}
\end{figure*}

Figure \ref{fig:sspjmpdf} shows the joint-marginalized PDFs when the assumed value matches the true value (the rest are in Appendix \ref{appxmockdata}). Effectively, this is a visualization of how well parameters can be recovered given a typical artificial star test uncertainty profile.

Using our core overshooting grid as the back-end stellar evolution models to \match, the $\asteca$-derived cluster members as input photometry, and the artificial star tests to account for photometric uncertainty and completeness, we evaluate Equation \ref{eqn:starprob} using \match\ or iterating calls to \match\ such that all combinations of  parameters listed in  Table \ref{tab:search} are searched.

\section{Results and Discussion}
\label{sec:analysis}

\subsection{Cluster Parameters With The Canonical PARSEC Model}
Before exploring the effects of uncertain core overshooting strength,  it is useful to understand the cluster parameter uncertainties  and their correlations while assuming the canonical PARSEC value of  \lambdac=0.50 \hp.  In effect, this is  a robust means to derive the cluster parameters if we were certain the most likely value of \lambdac\ was indeed 0.50 \hp. 

\begin{figure*}
\includegraphics[width=\textwidth]{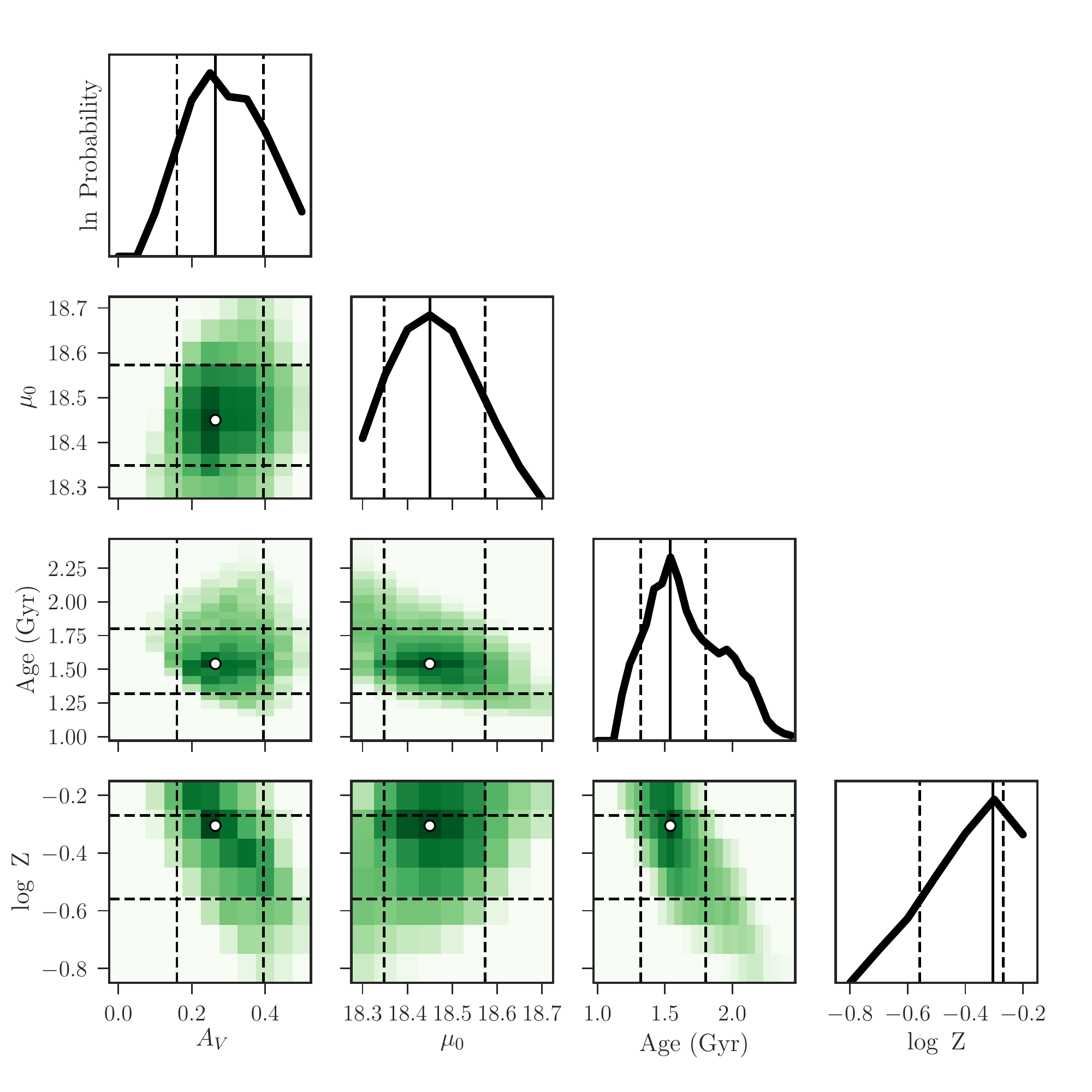}
\caption{Joint-marginalized PDFs derived with the canonical value of \lambdac=0.50 \hp\ for NGC~1795. Dashed lines mark the 16th and 84th quartiles of the polynomial fit to the distribution and solid vertical black lines on the diagonal panels mark the maximum posterior probability. The rest of the clusters are shown in in Appendix \ref{appxovis5}}\label{fig:ov5jmpdf}
\end{figure*}

Figure \ref{fig:ov5jmpdf} shows the joint marginalized PDFs for each cluster assuming \lambdac=0.50 \hp and the maximum posterior probabilities with the 16th and 84th quartiles of the distributions are listed in Table \ref{tab:ovis5results}. The most likely parameters agree reasonably well with previous work (see Table \ref{tab:lit}) given the different stellar models, fixed parameters, and fitting methods. The best agreement between our derivation and that in the literature is \citet{Goudfrooij2014}. Many of the derived values agree to within \citeauthor{Goudfrooij2014}'s reported uncertainties and most derived values agree to within our more conservative $\sim1 \sigma$ constraints from the PDFs.

\begin{deluxetable*}{lllll}
\tablecaption{Most Likely Cluster Parameters Given the PARSEC Model and canonical value of \lambdac=0.50 \hp}
\tablecolumns{11}
\tablewidth{0pt}
\tablehead{
\colhead{Cluster} &
\colhead{$A_V$} &
\colhead{$\mu_0$} &
\colhead{Age (Gyr)} &
\colhead{Z}
}
\startdata
HODGE~2 & $0.14^{+0.08}_{-0.16}$ &  $18.40^{+0.06}_{-0.15}$ &  $1.305^{+0.167}_{-0.179}$ &  $0.007^{+0.001}_{-0.002}$\\
NGC~1718& $0.55^{+0.15}_{-0.13}$ &  $18.65^{+0.13}_{-0.08}$ &  $1.833^{+0.274}_{-0.372}$ &  $0.006^{+0.002}_{-0.003}$\\
NGC~2203& $0.19^{+0.11}_{-0.19}$ &  $18.50^{+0.11}_{-0.10}$ &  $1.784^{+0.527}_{-0.417}$ &  $0.006^{+0.002}_{-0.003}$\\
NGC~2213& $0.15^{+0.07}_{-0.25}$ &  $18.50^{+0.11}_{-0.09}$ &  $1.591^{+0.381}_{-0.561}$ &  $0.008^{+0.001}_{-0.004}$\\
NGC~1644& $0.05^{+0.00}_{-0.27}$ &  $18.51^{+0.10}_{-0.09}$ &  $1.504^{+0.243}_{-0.340}$ &  $0.008^{+0.000}_{-0.004}$\\
NGC~1795& $0.26^{+0.10}_{-0.13}$ &  $18.45^{+0.10}_{-0.12}$ &  $1.541^{+0.221}_{-0.262}$ &  $0.008^{+0.001}_{-0.003}$\\
\enddata
\tablecomments{Most likely cluster parameters listed are the maximum posterior probability, given our priors and assuming \lambdac=0.50. Conservative uncertainties listed are the 16 and 84 percentiles of a polynomial fit to the posterior distributions (See Figures \ref{fig:ov5mpdf} and \ref{fig:ov5jmpdf}).}
\label{tab:ovis5results}
\end{deluxetable*}

\begin{figure*}
\includegraphics[width=\textwidth]{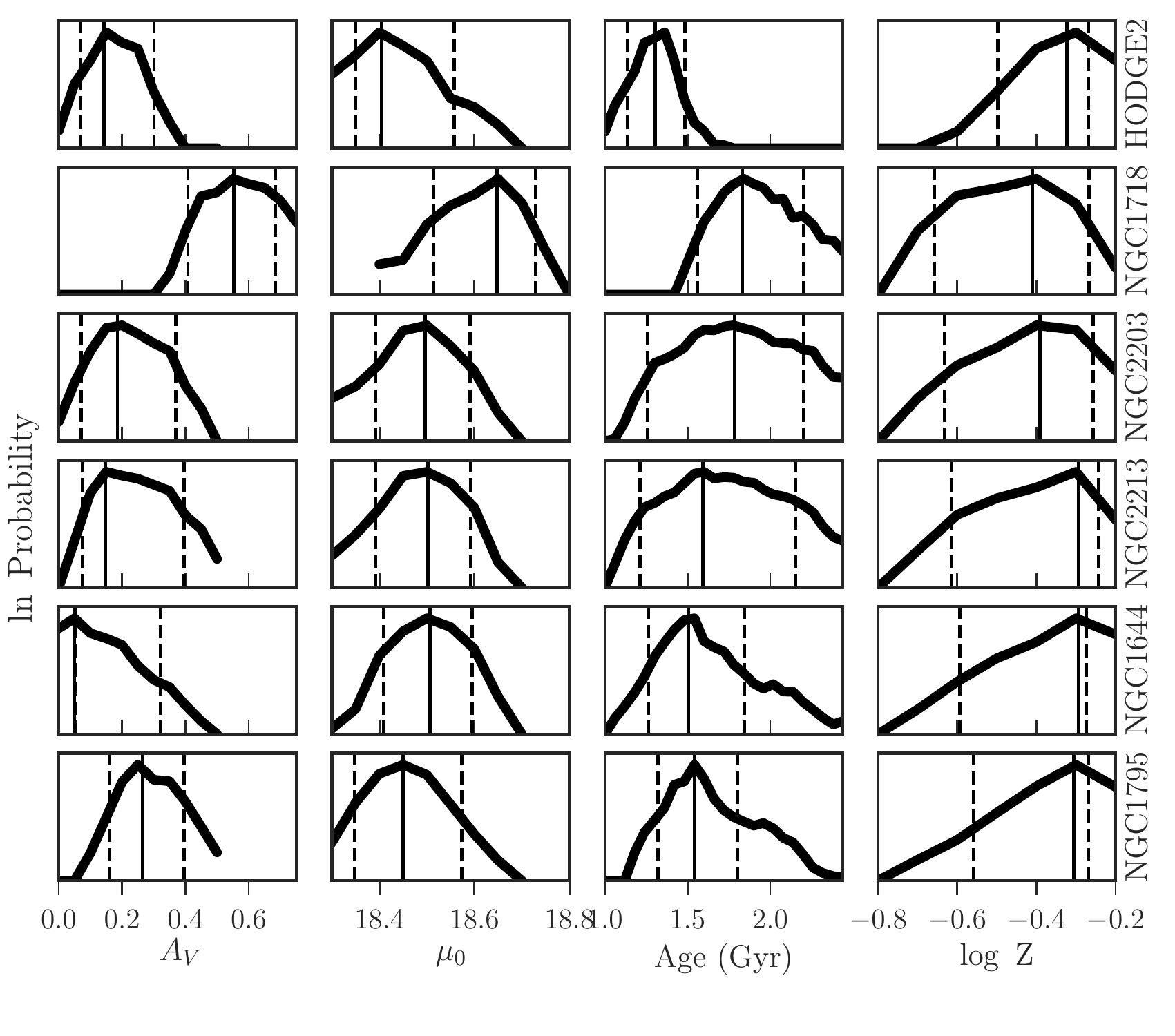}
\caption{Marginalized PDFs derived with the canonical value of \lambdac=0.50. Dashed lines mark the 16th and 84th quartiles of the polynomial fit to the distribution and solid vertical black lines mark the maximum posterior probability. Joint-marginalized PDFs are shown in Figure \ref{fig:ov5jmpdf} and Appendix \ref{appxovis5}.}\label{fig:ov5mpdf}
\end{figure*}

The main disagreements between \citet{Goudfrooij2014} and this work are in the fitting of NGC~1718, NGC~2203, and NGC~2213. Parameter differences for NGC~1718 and NGC~2203 are driven in part by the metallicity since \citeauthor{Goudfrooij2014} select the best fitting isochrone at either Z=0.008 or Z=0.006 and do not test intermediate values. For NGC~2213 the most likely metallicities agree, but we derive a distance $\sim0.7\%$ farther \citep[but closer to the mean LMC distance modulus of $18.49\pm0.09$][]{deGrijs2014} and a most probable age $\sim100$ Myr younger. We find the distance to NGC~1718 $\sim1\%$ farther than \citet{Goudfrooij2014} and $\sim0.7\%$ closer than \citet{Kerber2007}, though beyond the mean LMC distance modulus. 

Perhaps the most important aspect underlying the disagreements in cluster measurements are that these three clusters have eMSTOs. Determining the exact MSTO may be method dependent, especially for isochrone fitting. In other words, the nearly equal probable age over a span of ages evident in the PDFs make recovering the exact parameters difficult with the standard methods of isochrone fitting. The age-eMSTO connection reported in \citeauthor{Goudfrooij2014} and others' work is recovered in the relatively extended widths of the marginalized age PDFs. Since the morphology of the MSTO is the clearest signal on a CMD of the underlying population age, a spread MSTO age would certainly manifest as a spread in color and magnitude around the MSTO. However, we refrain from commenting on the cause of the eMSTO until fully rotational models are included within this framework.

Regardless of level of agreement between derived parameters, the joint marginalized PDFs (Figure \ref{fig:ov5jmpdf} and Figures in Appendix \ref{appxovis5}) reveal obvious correlations beyond the well known age-metallicity relationship. Metallicity appears correlated with extinction,  distance, and age for each cluster.  At the very least, these findings should give hesitation to heavily weighting parameters reported from isochrone fitting methods that fix values before attempting to constrain other parameters.

\subsection{Cluster Parameters Varying \Lambdac}
\begin{figure*}
\includegraphics[width=\textwidth]{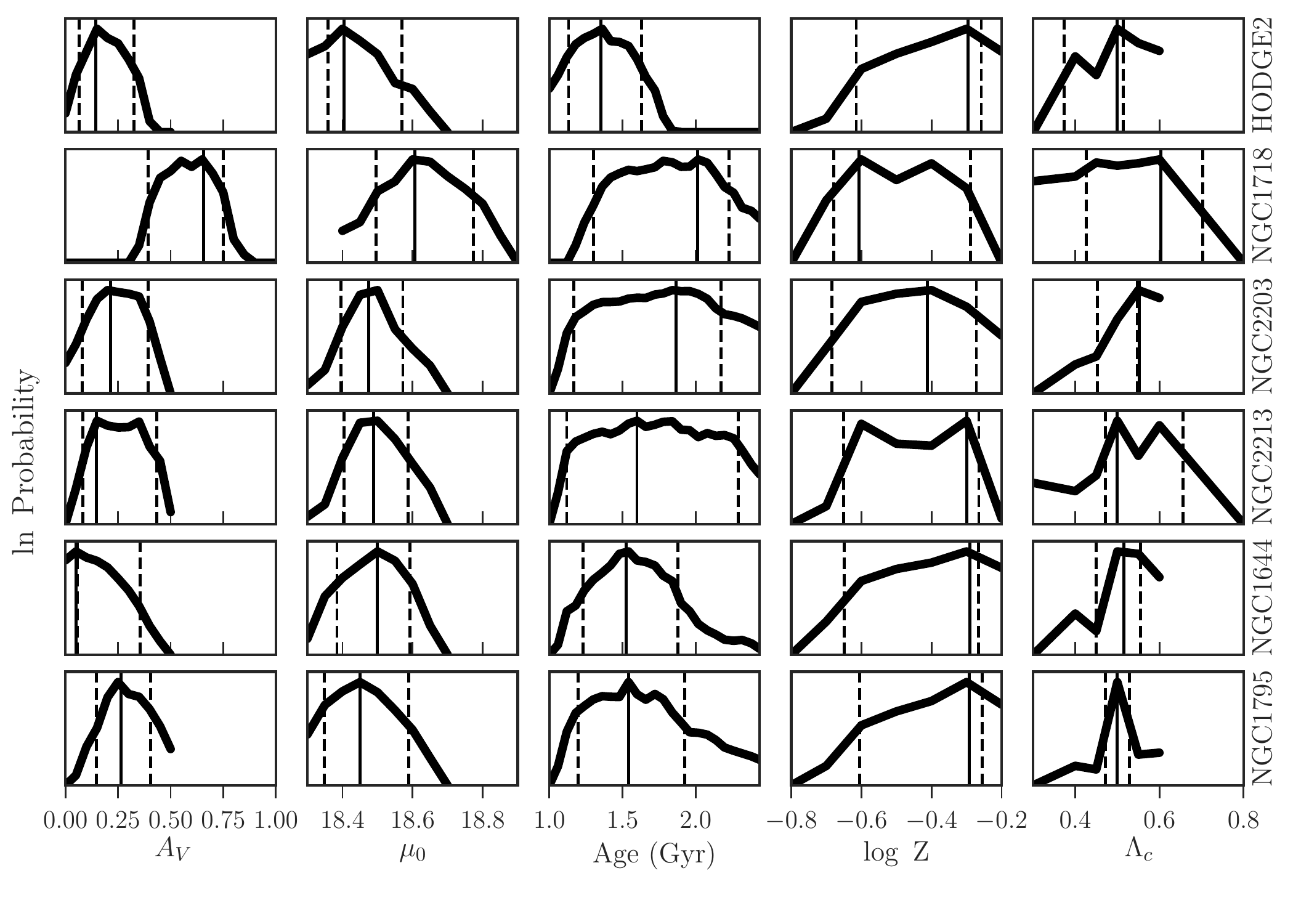}
\caption{Marginalized PDFs on observational and physical parameters for each cluster. Dashed lines mark the 16th and 84th quartiles of the polynomial fit to the distribution and solid vertical black lines mark the maximum posterior probability. Joint-marginalized posteriors are shown in Figure \ref{fig:margpdf} and Appendix \ref{appx}.}\label{fig:combo_margpdf}
\end{figure*}

Relaxing the core overshooting strength prior has the effect of spreading all the PDFs, though no significant changes are seen in the maximum posterior probabilities (see Figure \ref{fig:combo_margpdf}) with the exception of NGC~1718 which dropped 180 Myr in age and -0.2 dex in metallicity in order to land on its best fit core overshooting value of $\lambdac=0.6\ \hp$. The PDFs are also more complex compared to those in Figure \ref{fig:ov5jmpdf}. Core overshooting strength has a complex effect on CMDs, and some values of \lambdac\ seem to align well with different values of metallicity (e.g., NGC~2213). Asymmetric or lopsided PDFs are not signs of poor data quality or unreliable models, they are only signs that Gaussian and perhaps other functional approximations will likely inadequately describe the distribution.

The marginalized PDFs of convective core overshooting vary dramatically from cluster to cluster. For example, Figure \ref{fig:margpdf} shows the joint marginalized PDFs for NGC~1795.  There are clear peaks in each PDF denoting the maximum posterior probabilities which are listed in Table \ref{tab:results}.  The general trends in correlations between parameters in the top four rows are very similar to in Figure \ref{fig:ov5jmpdf} when \lambdac\ was fixed to the PARSEC canonical values.  However, in the bottom row there are now correlations between \lambdac\ and other cluster parameters. The most apparent is the correlation between core overshooting and age. Next, there are slight correlations with \lambdac\ and distance and \lambdac\ and metallicity (the \lambdac-metallicity correlation is built into the PARSEC models). These correlations apparent in the joint marginalized PDF of NGC~1795 are also seen in all other clusters (see Appendix \ref{appx}).

One of many robust ways of discerning if the effect one measures is actually due to the parameter in question is by removing the parameter and re-running the analysis, and still understanding the results. By presenting our PDFs pedagogically, that is without varying core overshooting, and then by varying core overshooting, we have effectively done the necessary reliability test but in reverse.  All the changes in the PDFs introduced by allowing core overshooting to vary are expected from the preceding discussion on the effects of core overshooting. For a couple clusters (e.g., NGC2213), a higher value of overshooting  with a lower value of metallicity fit nearly  as well as the  most probable values with canonical overshooting.

\begin{figure*}
\includegraphics[width=\textwidth]{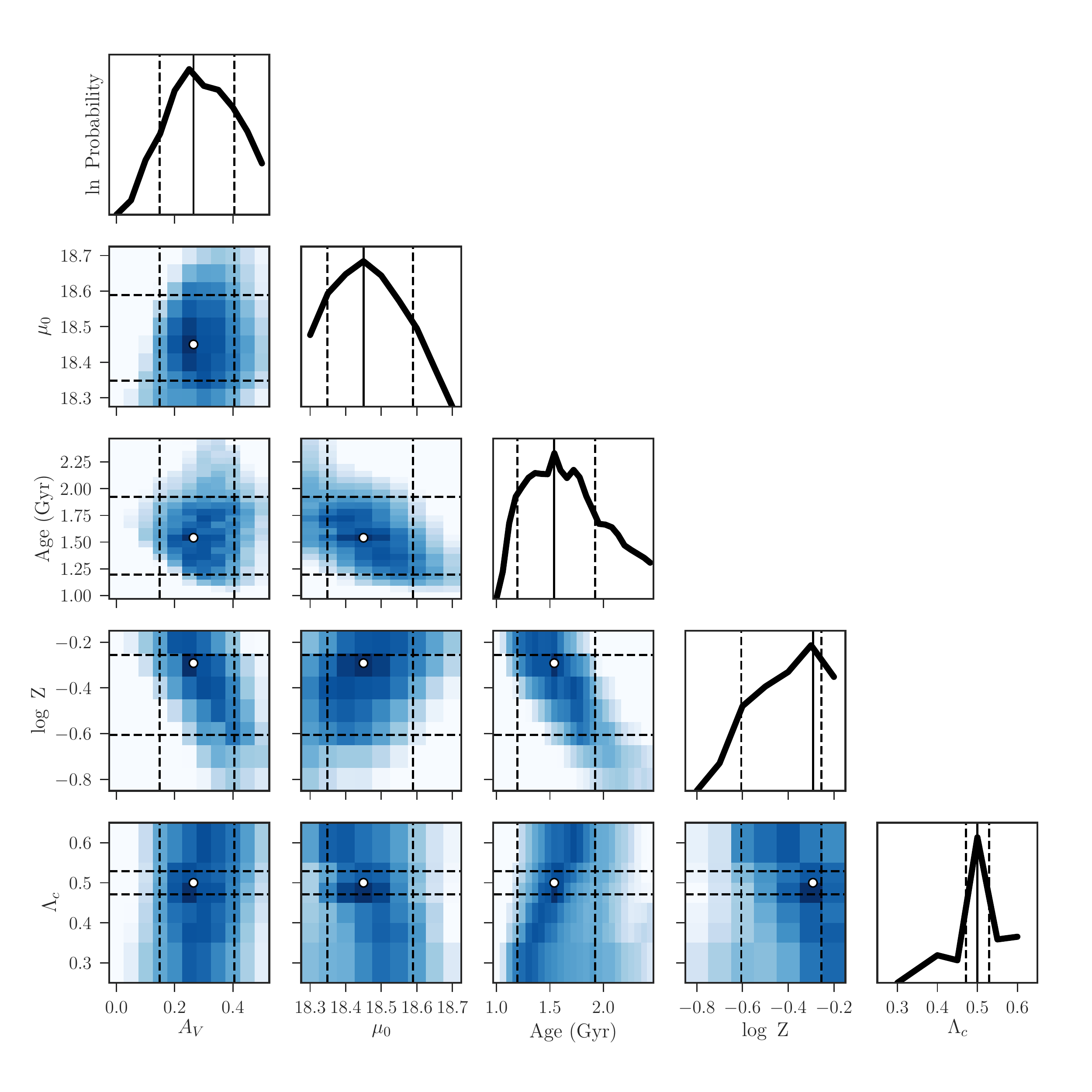}
\caption{Joint-marginalized PDFs for NGC~1795. Dashed lines mark the 16th and 84th quartiles of the polynomial fit to the distribution and solid vertical black lines on the diagonal panels mark the maximum posterior probability. The diagonal panels for each cluster are the marginalized posteriors, shown together in Figure \ref{fig:combo_margpdf}. The joint-marginalized PDFs for the remaining clusters are in Appendix \ref{appx}.}\label{fig:margpdf}
\end{figure*}

As discussed in Section \ref{sec:convec}, the emerging trend in recent studies focused on individual (or binary) stars is toward core overshooting strength increasing with increasing stellar mass (or age). It is interesting in this context that we find a strong correlation between age and \lambdac\ in exactly the same direction over an age range of $\sim$1-2.5 Gyr (the exact age limits depend on the cluster).  In other words, a younger age will be derived from a stellar population with true core overshooting strength that is lower than the model (see also, the bottom two panels of Figure \ref{fig:sspmpdf}). Researchers constraining core overshooting (or perhaps any stellar evolutionary parameter) should be vigilant of possible degeneracies and their implications on their results.

We combined the marginalized PDFs of core overshooting strength (by summing the log likelihoods of all 6 clusters), and evaluated the resulting maximum posterior probability, the 16th, and the 84th quartiles. Given the PARSEC models, we find the most likely value of core overshooting for clusters with ages $\sim1.3-2.0$ Gyr is $\lambdac=0.500^{+0.016}_{-0.134} \hp$.  Our results fit within the previous work  of \citet{Girardi2009}, but is a slightly more efficient value than  expected in the relationships presented by \citet{Claret2016}.

\section{Conclusions}\label{sec:conclusions}
Convection is an important but uncertain aspect of stellar evolution. We show that uncertainty in the strength of core overshooting can result in $\sim150$ Myr uncertainty in core burning lifetimes for stars with mass $\sim1-2$\msun.  This timescale is nearly as long as the expected SF duration invoked as an explanation of extended MSTOs in the MCs, the lifetimes of massive Helium burning stars in nearby dwarf galaxies, and the lifetimes of important, but short lived stellar phases like the TP-AGB. 

We have introduced a robust method to constrain uncertain stellar evolutionary parameters and applied the method to simultaneously fit foreground extinction, distance, age, metallicty, and the strength of core overshooting using 6 LMC clusters with a narrow range of previously reported ages ($1.30-2.04$ Gyr). We report the most likely cluster parameters as well as the correlations between the parameters. We show several strong correlations, even when fixing \lambdac to the canonical PARSEC value. Metallicity appears correlated with extinction,  distance, and age for each cluster. When varying \lambdac, we find a strong correlation with increasing core overshooting strength and increasing age, mirroring trends reported in the literature. 

This study is a first step in systematically constraining uncertain aspects of stellar evolution using MC clusters. We expected clusters within the range of $\sim1.5$ Gyr MSTO would have core overshooting strength at roughly $0.4\ \hp<\lambdac<0.5\ \hp$. Our findings on the most likely values were expected, however, the complex shape of the PDFs and the strength of the degeneracies between $\lambdac$ and age were perhaps surprising.  We will apply this fitting method to MC clusters at  various literature-derived ages to further test whether or not core overshooting does in fact increase with increasing mass.

We will explore other means to investigate the relationship between core overshooting and age. For example, we will to try to break the correlation by imposing stronger prior distributions. One way to do this would be to use Milky Way open clusters that have independently-derived ages, such as from white dwarf cooling sequences or gyrochronology \citep[e.g.,][]{Jeffery2011, Tremblay2014, Barnes2007}. 

Applying more independent measurements to constrain prior distributions should also tighten the PDFs. Such measurements would be especially beneficial for metallicity, given its correlations with other cluster parameters, and the multiple-peaked marginalized PDFs (see NGC~1718 and NGC~2213 in Figure \ref{fig:margpdf}). For example, including spectroscopically determined metallicities  of a sample of stars in the clusters would help further constrain \lambdac, or any other physical model, such as rotation. In light of the correlations found between the cluster parameters, we urge caution when using results from isochrone fitting methods that fix or adopt values before actually fitting.

\begin{deluxetable*}{llllll}
\tablecaption{Most Likely Cluster Parameters Given the PARSEC Model}
\tablecolumns{11}
\tablewidth{0pt}
\tablehead{
\colhead{Cluster} &
\colhead{$A_V$} &
\colhead{$\mu_0$} &
\colhead{Age (Gyr)} &
\colhead{Z} &
\colhead{$\lambdac$}
}
\startdata
HODGE~2 & $0.14^{+0.18}_{-0.08}$ &  $18.40^{+0.17}_{-0.04}$ &  $1.354^{+0.277}_{-0.222}$ &  $0.008^{+0.001}_{-0.004}$ &  $0.500^{+0.014}_{-0.127}$\\
NGC~1718& $0.66^{+0.09}_{-0.27}$ &  $18.61^{+0.17}_{-0.11}$ &  $2.015^{+0.213}_{-0.712}$ &  $0.004^{+0.004}_{-0.001}$ &  $0.603^{+0.100}_{-0.177}$\\
NGC~2203& $0.22^{+0.18}_{-0.14}$ &  $18.47^{+0.10}_{-0.07}$ &  $1.867^{+0.308}_{-0.699}$ &  $0.006^{+0.002}_{-0.003}$ &  $0.552^{+0.004}_{-0.100}$\\
NGC~2213& $0.15^{+0.29}_{-0.06}$ &  $18.49^{+0.10}_{-0.08}$ &  $1.599^{+0.693}_{-0.480}$ &  $0.008^{+0.001}_{-0.004}$ &  $0.500^{+0.157}_{-0.028}$\\
NGC~1644& $0.05^{+0.30}_{-0.01}$ &  $18.50^{+0.09}_{-0.11}$ &  $1.524^{+0.353}_{-0.294}$ &  $0.008^{+0.000}_{-0.004}$ &  $0.516^{+0.039}_{-0.067}$\\
NGC~1795& $0.27^{+0.14}_{-0.12}$ &  $18.45^{+0.14}_{-0.10}$ &  $1.541^{+0.383}_{-0.344}$ &  $0.008^{+0.001}_{-0.004}$ &  $0.500^{+0.029}_{-0.029}$\\
\enddata
\tablecomments{Most likely cluster parameters listed are the maximum posterior probability, given our priors. Conservative uncertainties listed are the 16 and 84 percentiles of the PDFs (See Figures \ref{fig:combo_margpdf} and \ref{fig:margpdf}).}
\label{tab:results}
\end{deluxetable*}

\clearpage
\appendix

\section{Joint Marginalized Posterior Distribution Functions of Synthetic Stellar Populations}\label{appxmockdata}

\begin{figure*}[!ht]
\includegraphics[width=\textwidth]{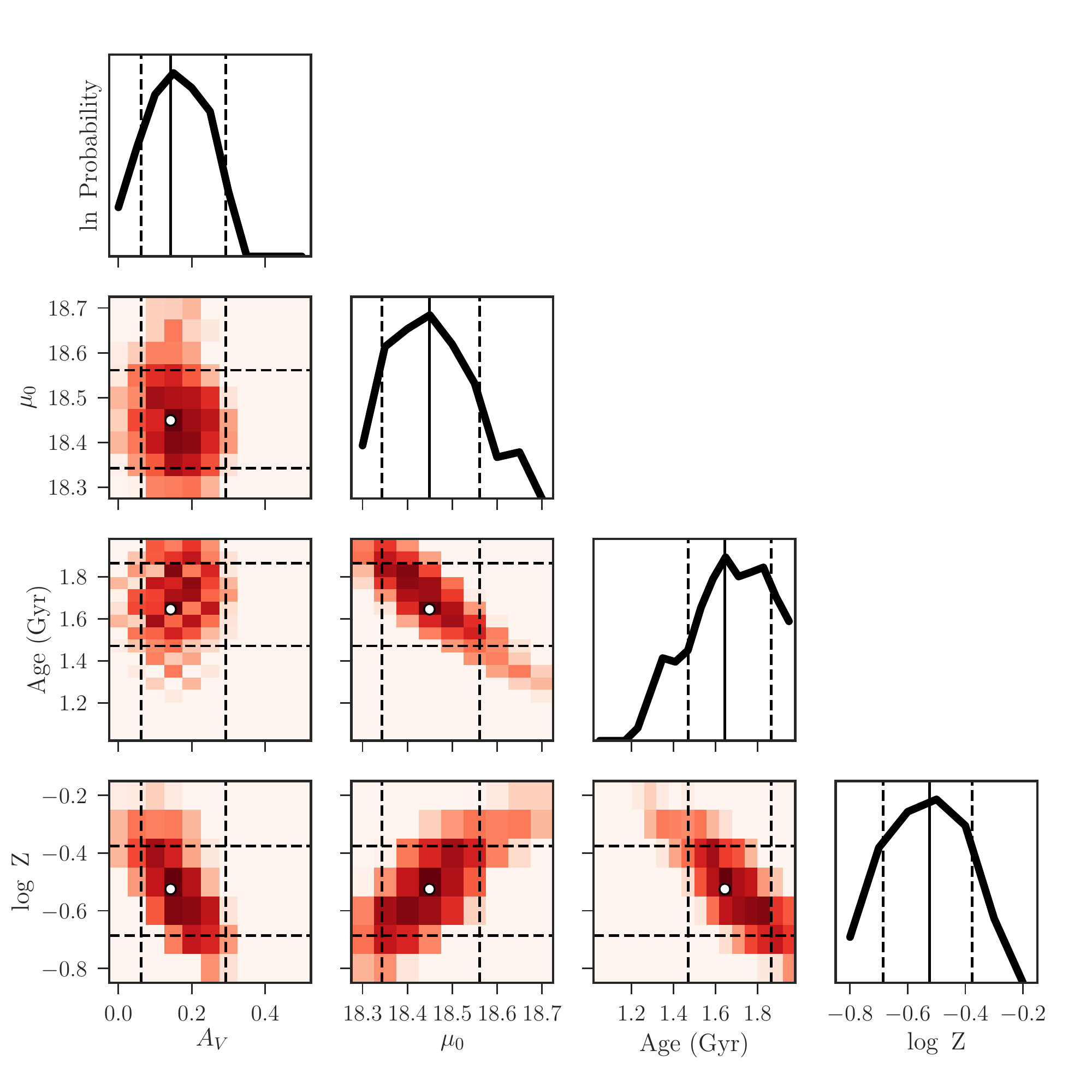}
\figurenum{\ref{fig:sspjmpdf}}
\caption{continued, with the stellar population calculated with \lambdac=0.3 \hp}
\end{figure*}

\begin{figure*}
\includegraphics[width=\textwidth]{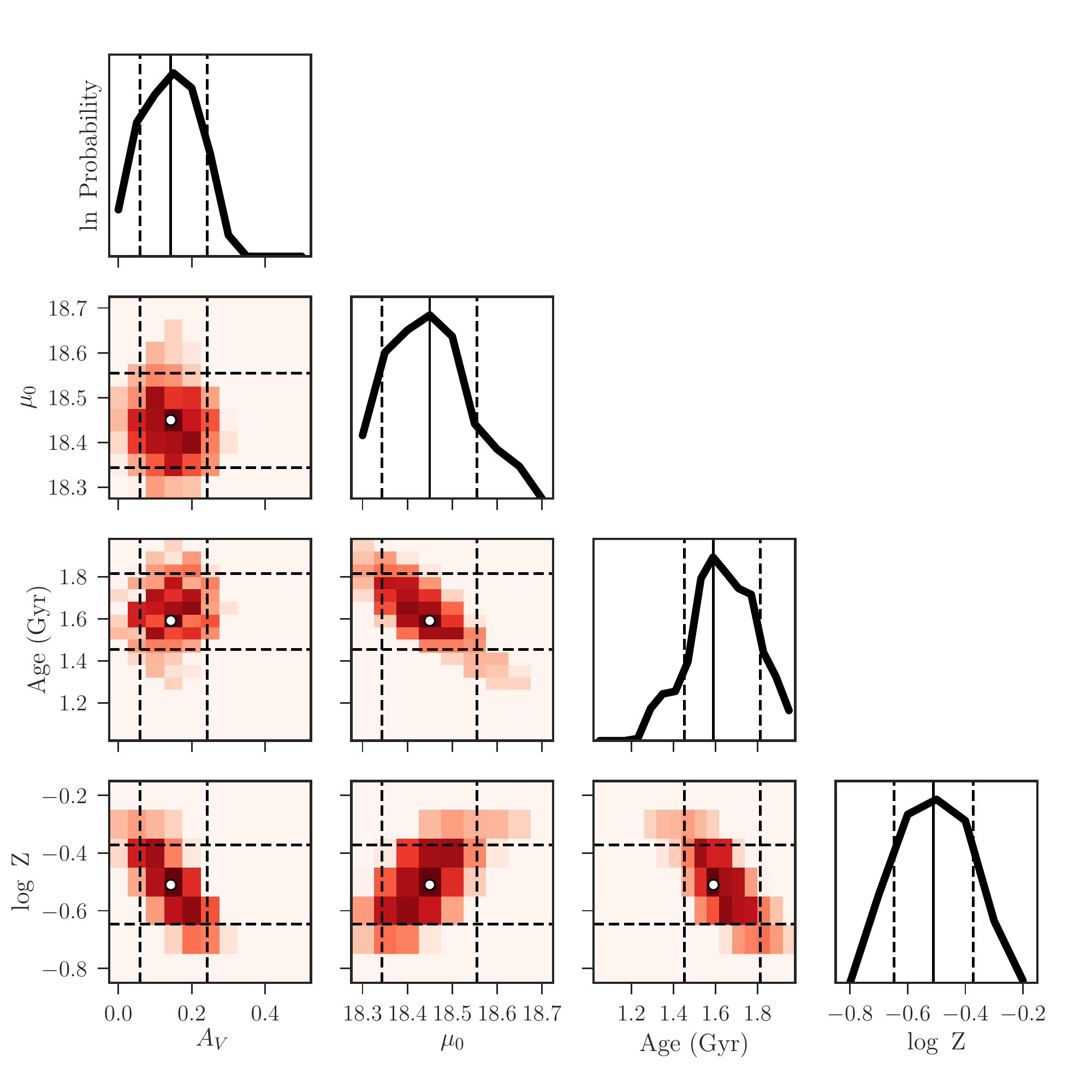}
\figurenum{\ref{fig:sspjmpdf}}
\caption{continued, with the stellar population calculated with \lambdac=0.4 \hp}
\end{figure*}

\begin{figure*}
\includegraphics[width=\textwidth]{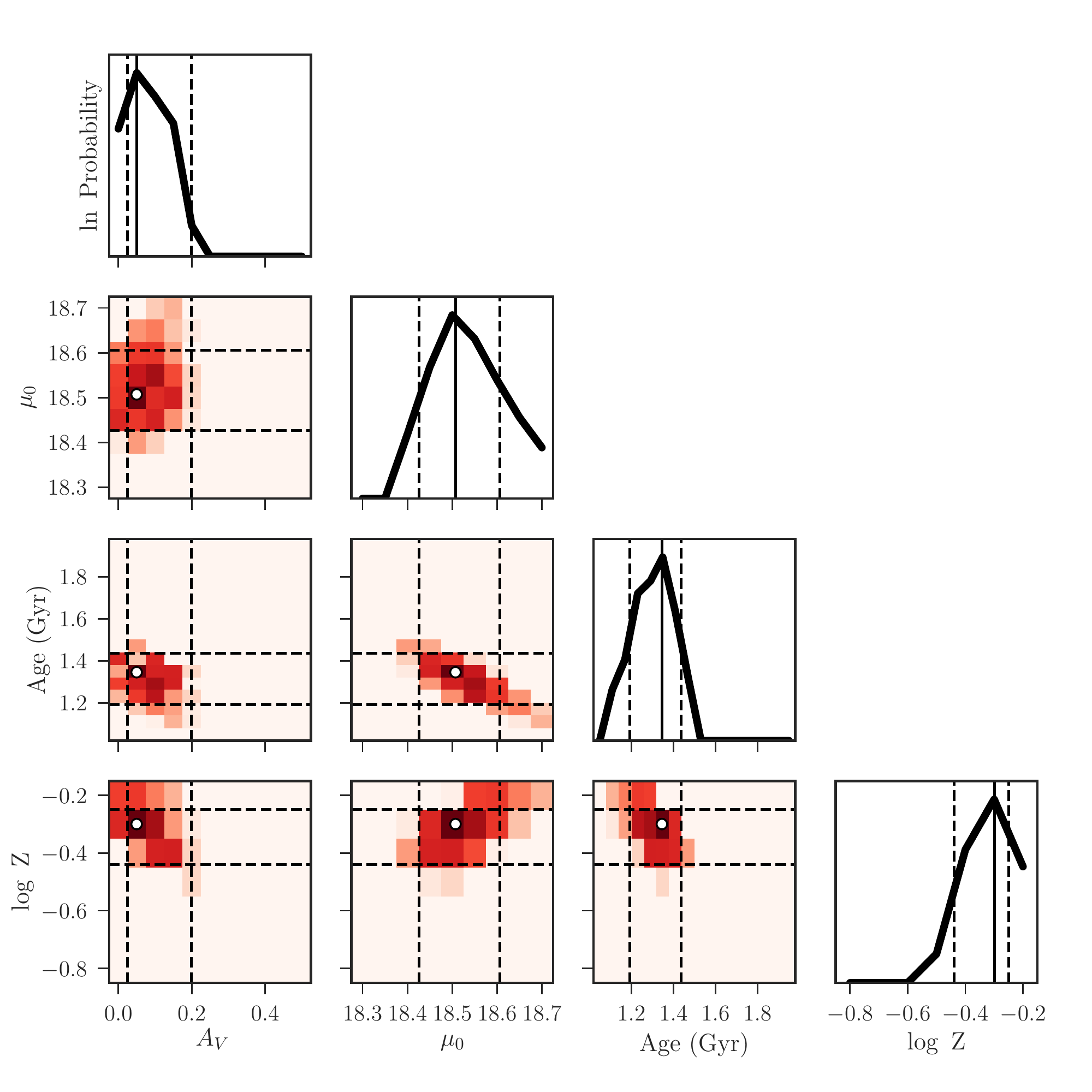}
\figurenum{\ref{fig:sspjmpdf}}
\caption{continued, with the stellar population calculated with \lambdac=0.6 \hp}
\end{figure*}

\clearpage
\section{Joint Marginalized Posterior Distribution Functions With Canonical \lambdac=0.50}\label{appxovis5}

\begin{figure*}
\includegraphics[width=\textwidth]{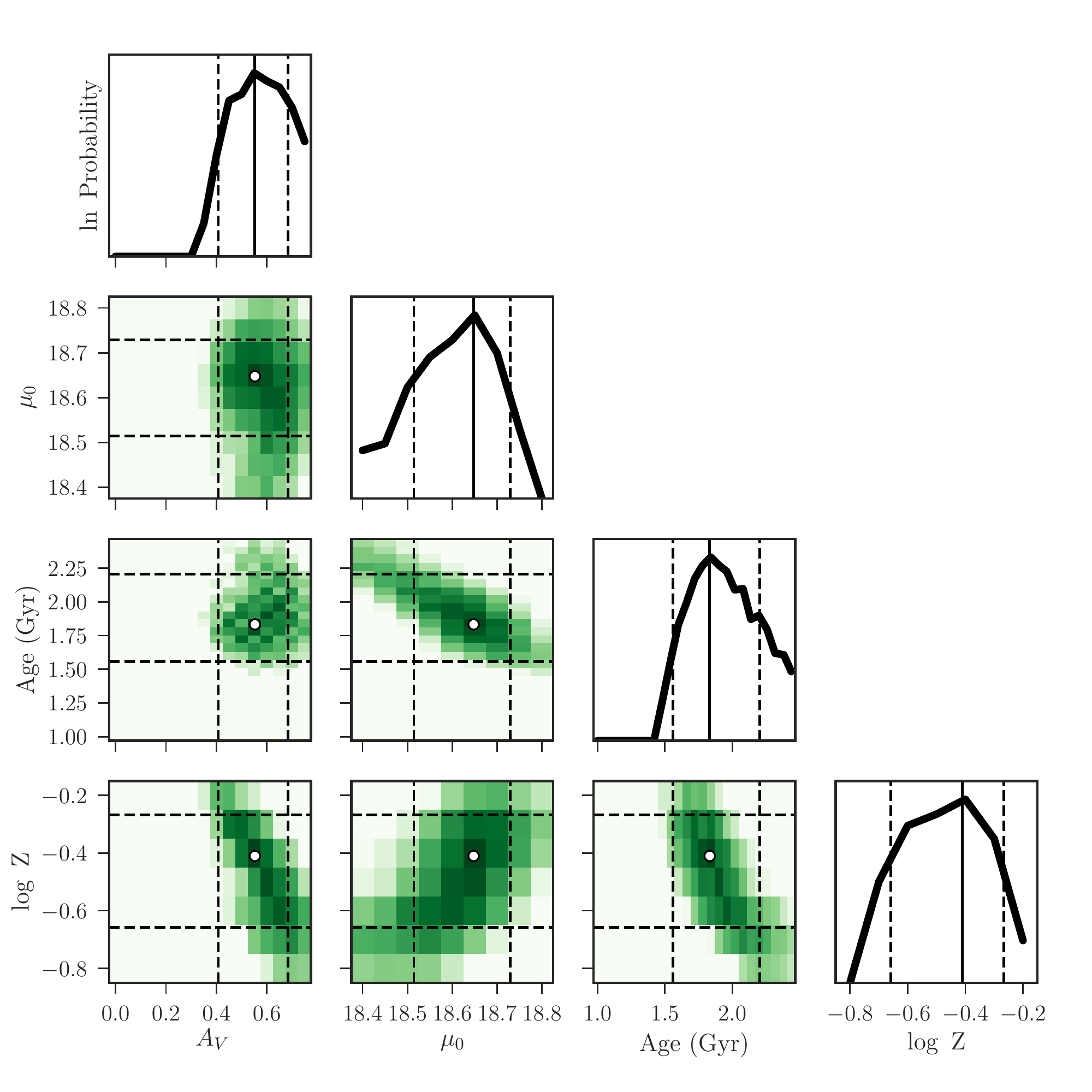}
\figurenum{\ref{fig:ov5jmpdf}}
\caption{continued, with NGC~1718}
\end{figure*}

\begin{figure*}
\includegraphics[width=\textwidth]{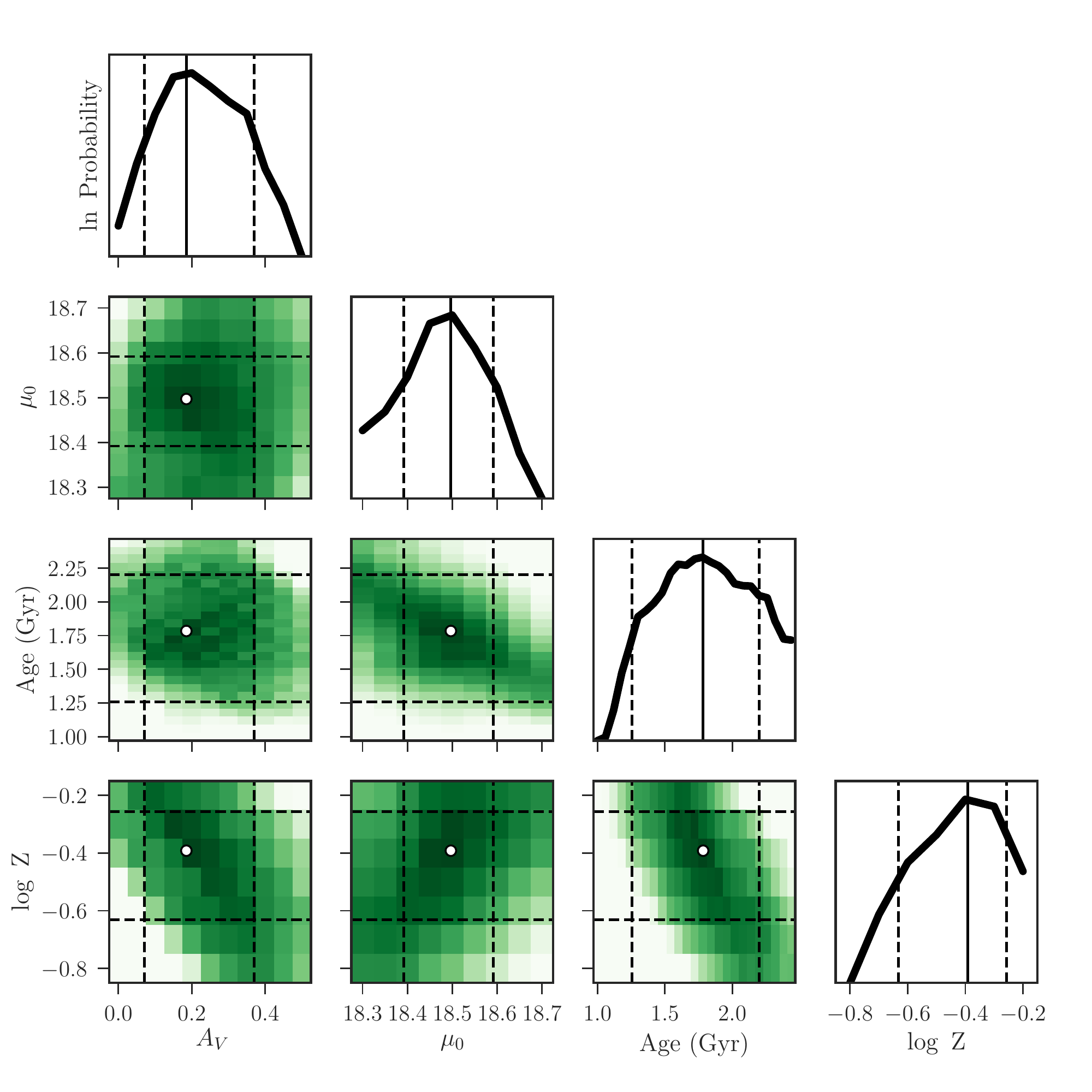}
\figurenum{\ref{fig:ov5jmpdf}}
\caption{continued, with NGC~2203}
\end{figure*}

\begin{figure*}
\includegraphics[width=\textwidth]{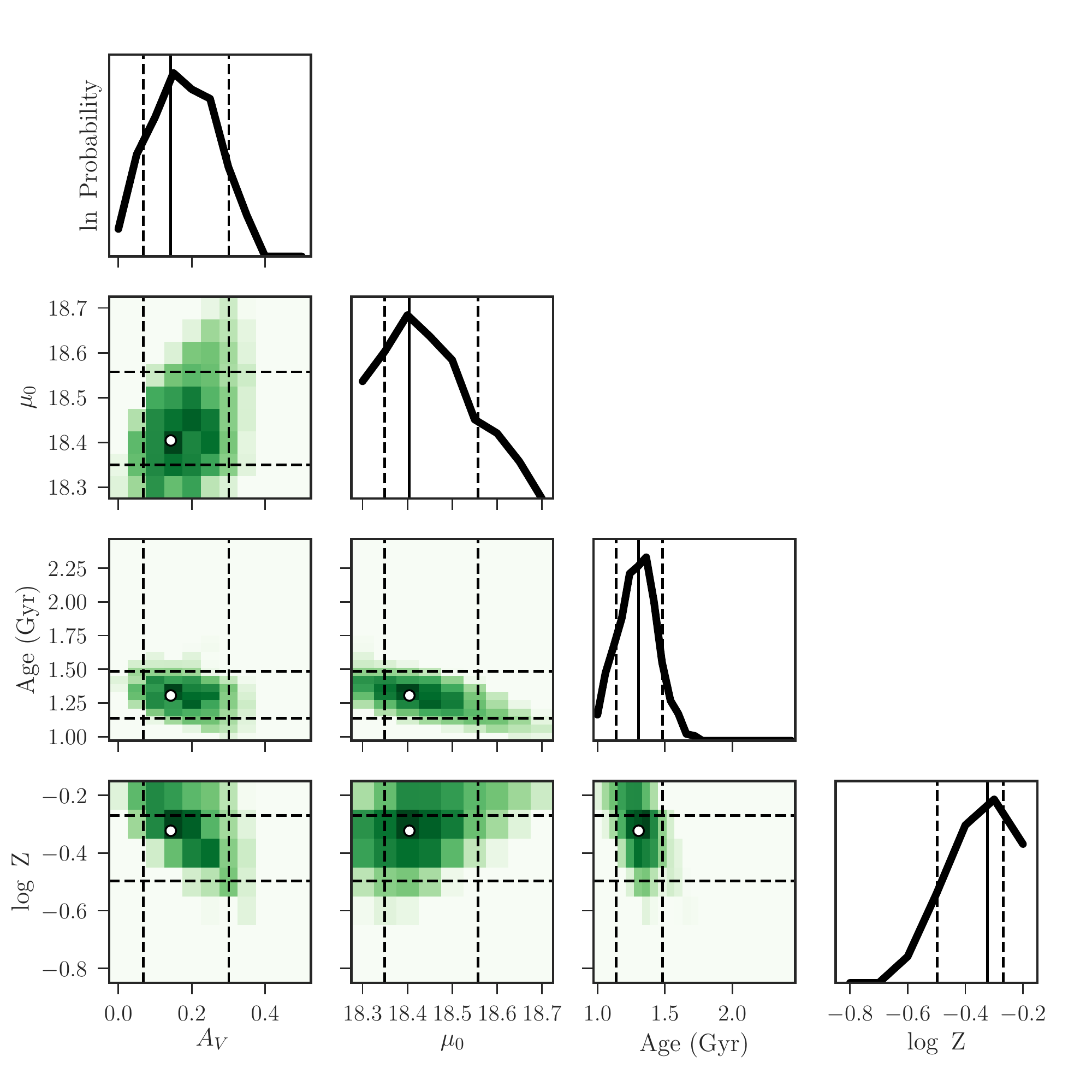}
\figurenum{\ref{fig:ov5jmpdf}}
\caption{continued, with HODGE~2}
\end{figure*}

\begin{figure*}
\includegraphics[width=\textwidth]{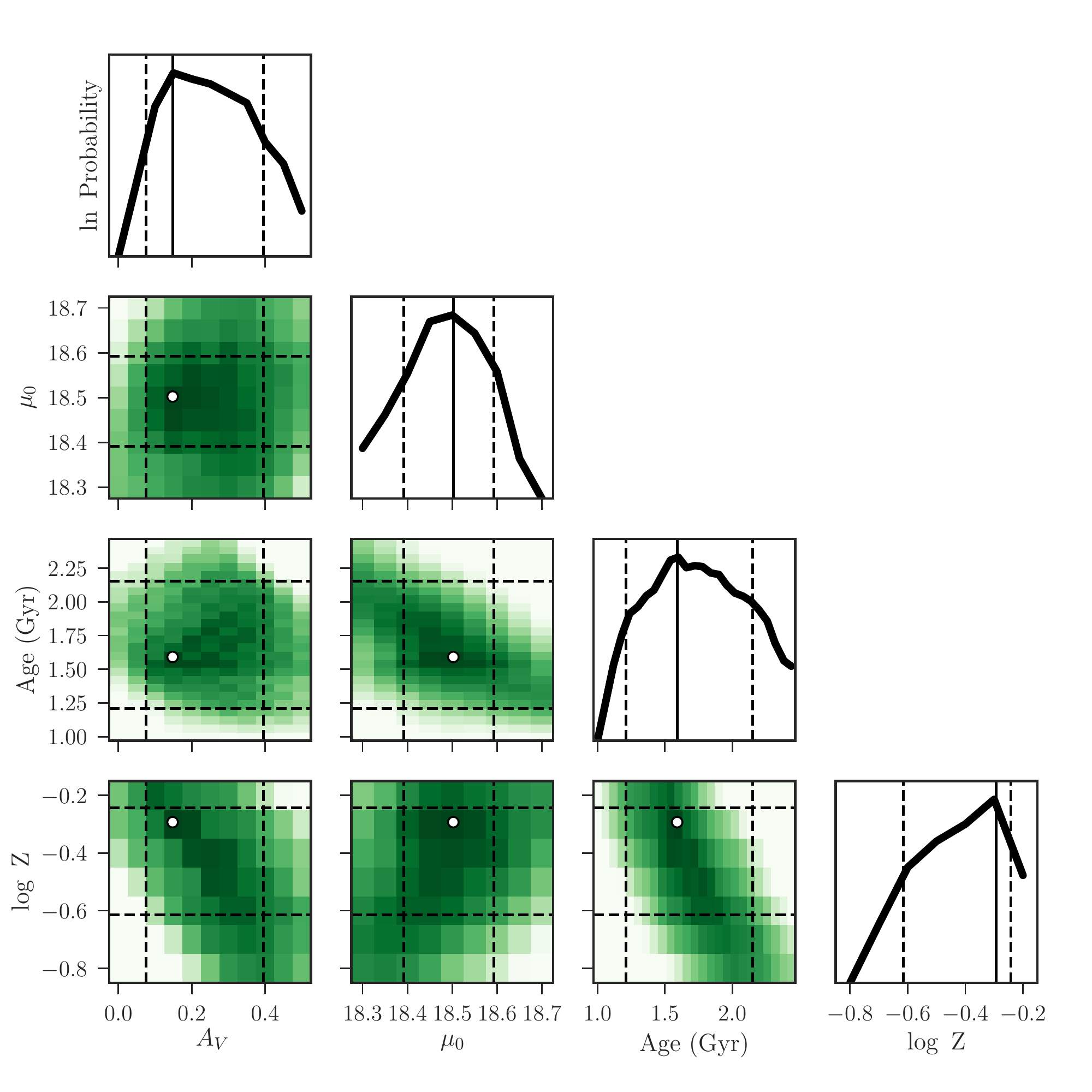}
\figurenum{\ref{fig:ov5jmpdf}}
\caption{continued, with NGC~2213}
\end{figure*}

\begin{figure*}
\includegraphics[width=\textwidth]{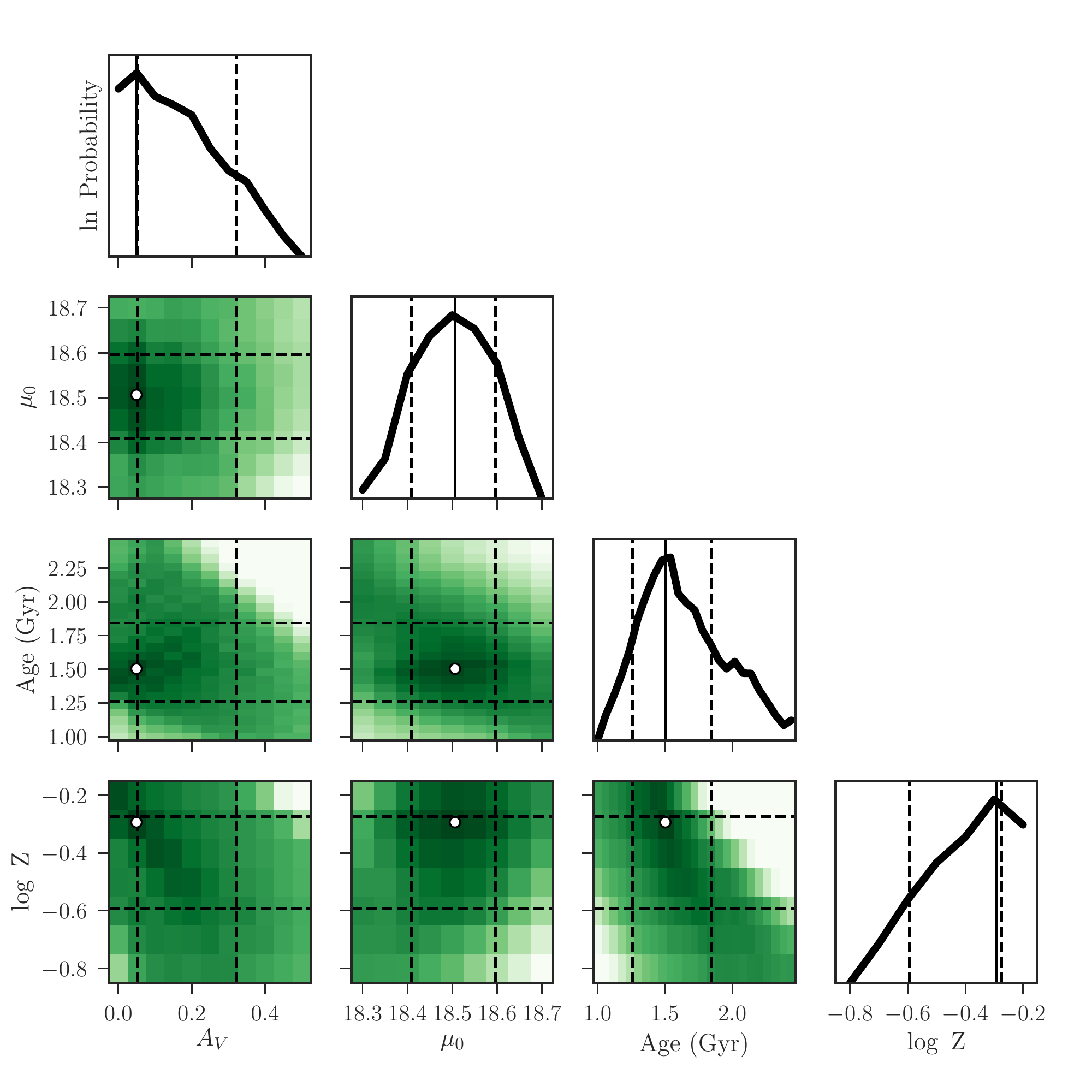}
\figurenum{\ref{fig:ov5jmpdf}}
\caption{continued, with NGC~1644}
\end{figure*}

\clearpage
\section{Joint Marginalized Posterior Distribution Functions}\label{appx}

\begin{figure*}[!ht]
\includegraphics[width=\textwidth]{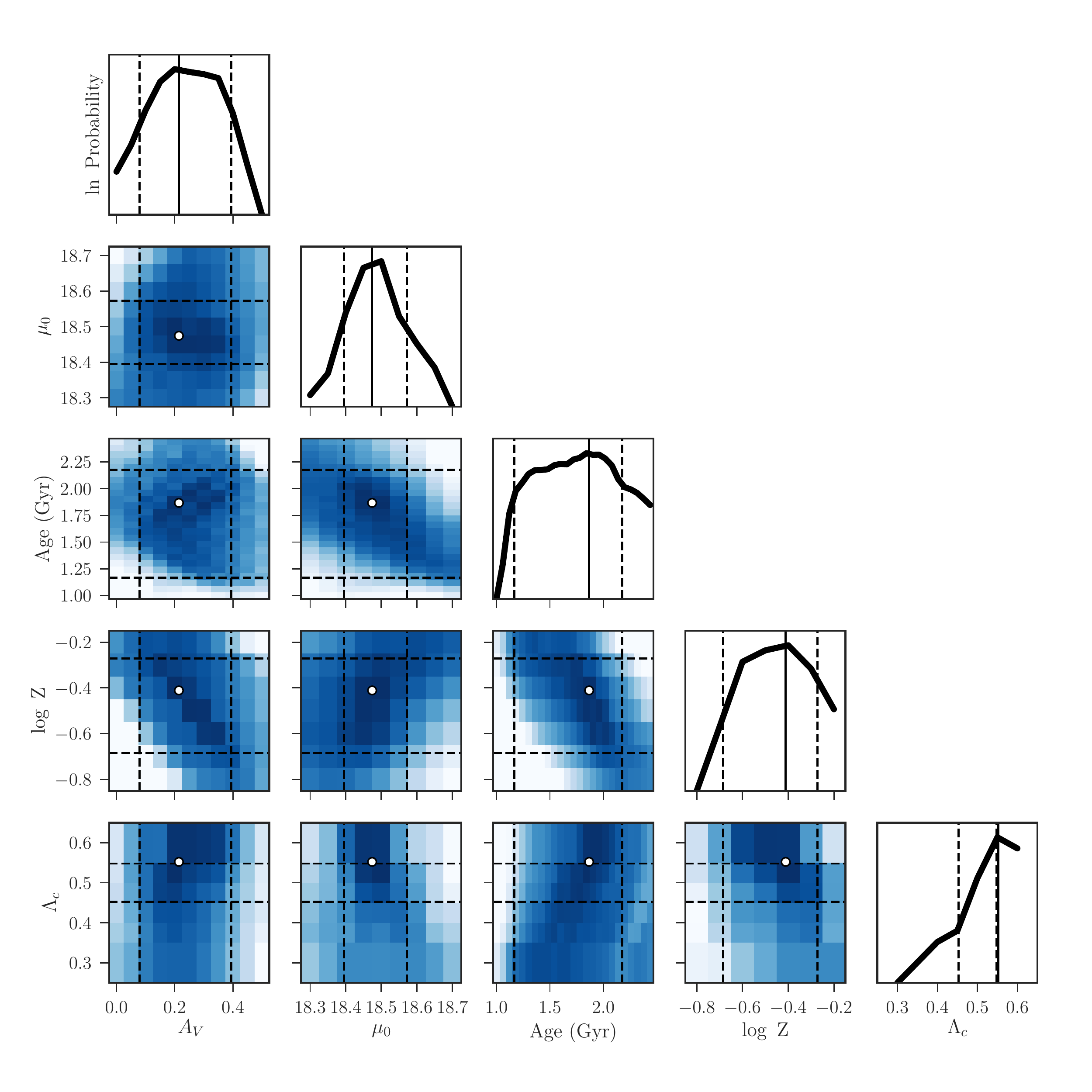}
\figurenum{\ref{fig:margpdf}}
\caption{continued, with NGC~2203}
\end{figure*}

\begin{figure*}
\includegraphics[width=\textwidth]{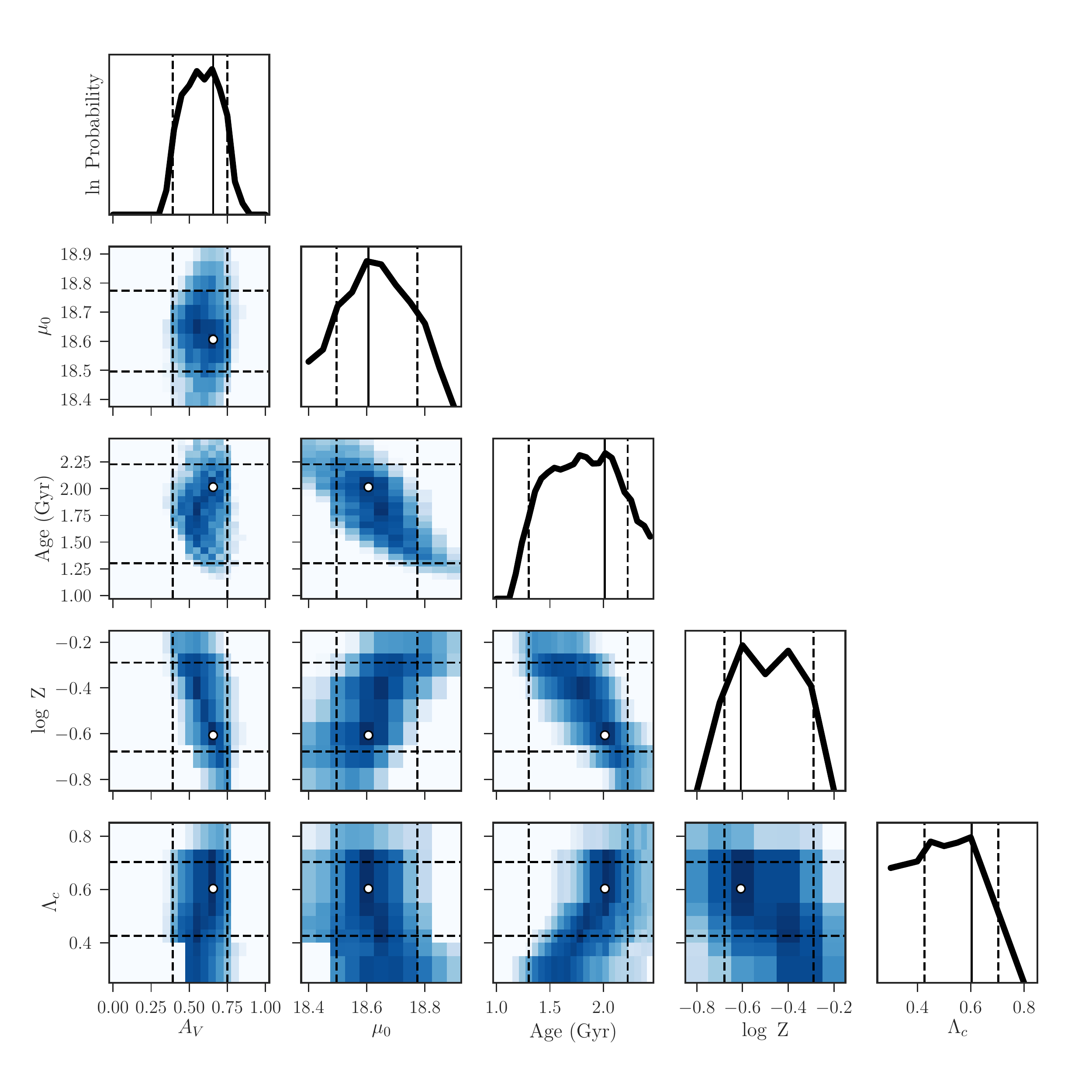}
\figurenum{\ref{fig:margpdf}} 
\caption{continued, with NGC~1718}
\end{figure*}

\begin{figure*}
\includegraphics[width=\textwidth]{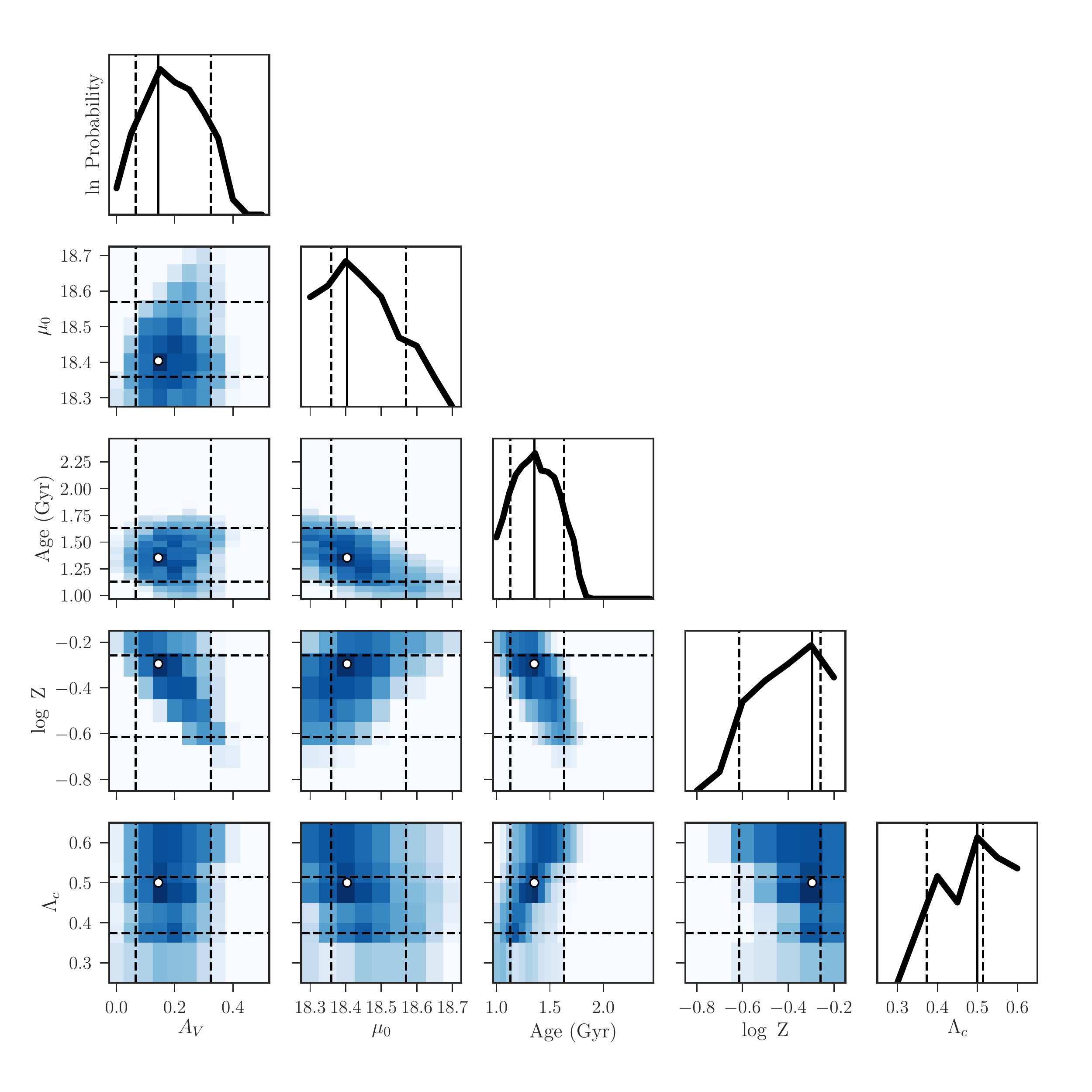}
\figurenum{\ref{fig:margpdf}}
\caption{continued, with HODGE~2}
\end{figure*}

\begin{figure*}
\includegraphics[width=\textwidth]{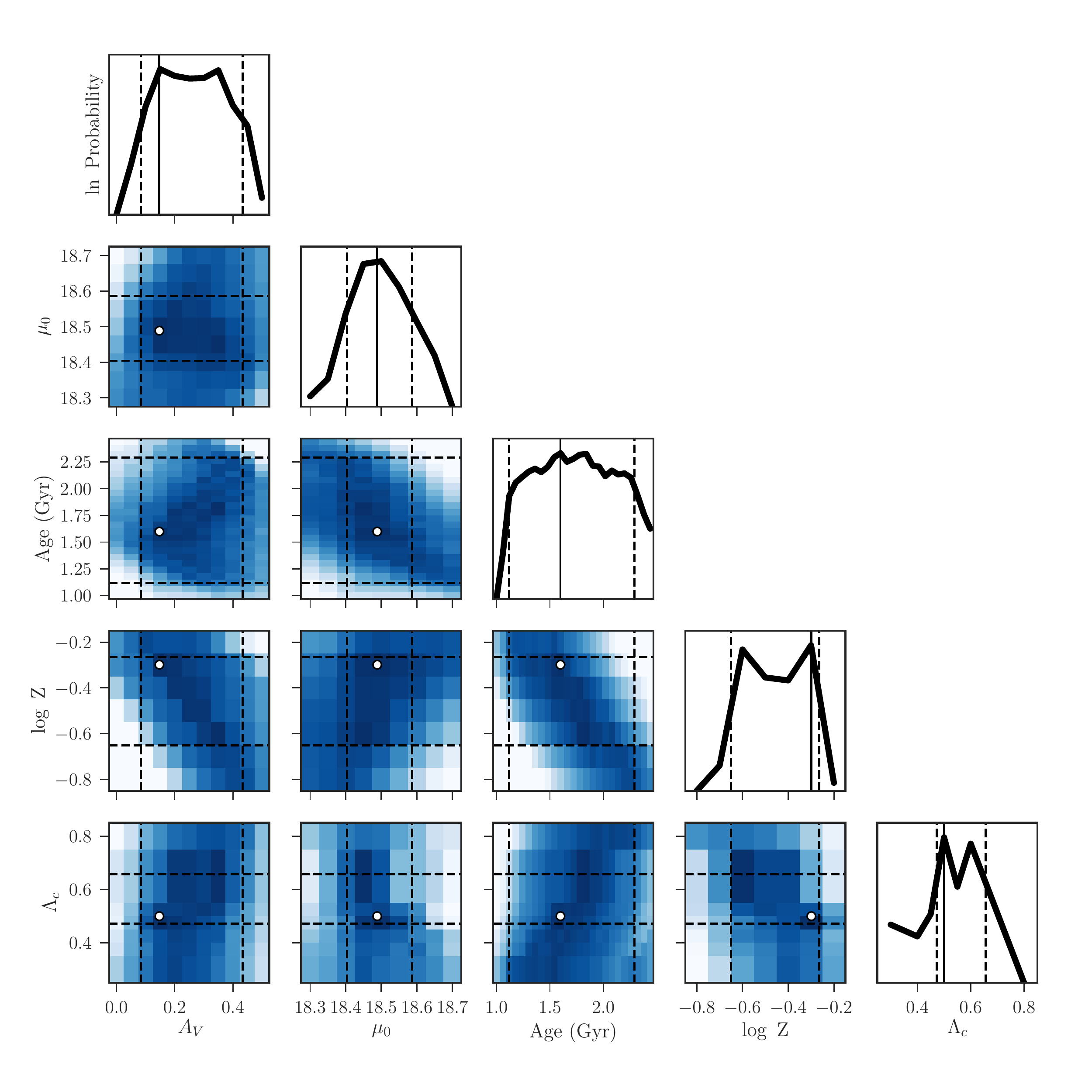}
\figurenum{\ref{fig:margpdf}} 
\caption{continued, with NGC~2213}
\end{figure*}

\begin{figure*}
\includegraphics[width=\textwidth]{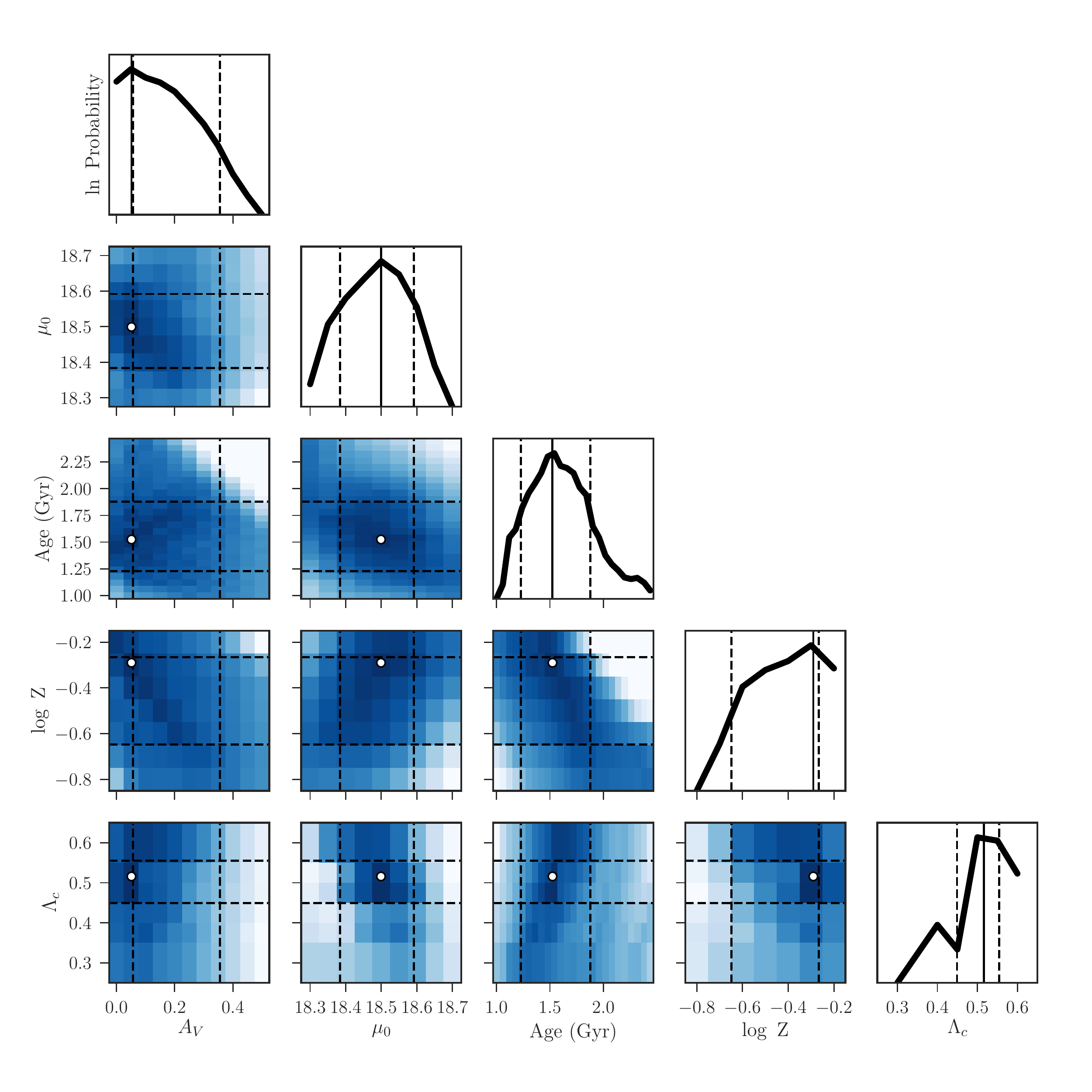}
\figurenum{\ref{fig:margpdf}}
\caption{continued, with NGC~1644}
\end{figure*}
\clearpage

\section{Uncertainties Across Stellar Modeling Groups}
\label{appx:othermodels}
We motivated this study with the statement that stellar evolution models are fundamental to nearly all studies in astrophysics and implied the importance of a quantitative understanding of uncertainties within stellar models. However, this study begs the question of what to do with vastly different predictions {\it across} stellar modeling groups. Stellar models (i.e., tracks or isochrones) are seldom published with any estimates of uncertainties, leaving researchers who use the models to fend for themselves (or assume infinite precision).  In one strategy, researchers have applied models from different stellar evolution groups and considered differences in predictions to be systematic uncertainties of stellar models \citep[e.g.,][]{Weisz2014,Dolphin2012}.  Until stellar evolution groups provide probabilistic tracks and isochrones to the community, this is probably the most reasonable means to interpret results from models that use different input physical assumptions.

However, from the perspective of a stellar evolutionist, differences between a stellar model from one group to another is not a source of uncertainty. In fact, the choices made in each group are very deliberate. For example, models rest on some Solar calibration to scale abundances heavier than He, but differ on the source of the calibration and therefore, the initial Solar metallicity. Models also differ in their treatment (i.e., applications of 1D approximations) of convection, applying different mixing length parameter values ($\alpha_{\rm MLT}$; which are also calibrated to a Solar model) as well as different treatments of convective overshooting, from MLT-like, \citep[e.g., YaPSI and Dartmouth;][]{Spada2017, Dotter2008} to a diffusion approximation \citep[e.g., MIST, Victoria-Regina][]{Choi2016, VandenBerg2006}. Still, each of the above listed modeling groups report the effective strength in core convective overshooting is $\lambdac\sim0.4\ \hp$.

\begin{figure*}
\includegraphics[width=0.5\textwidth]{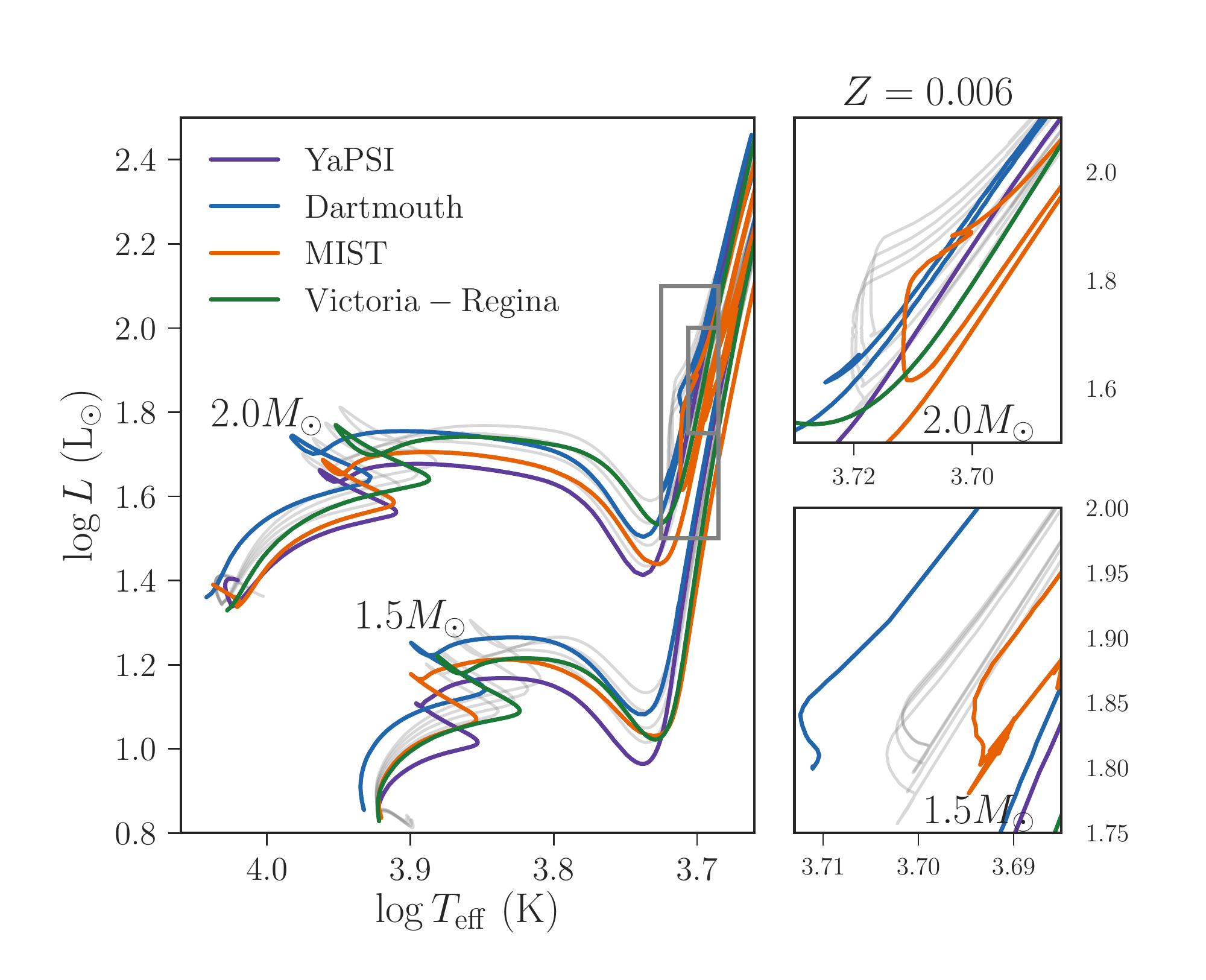}
\includegraphics[width=0.5\textwidth]{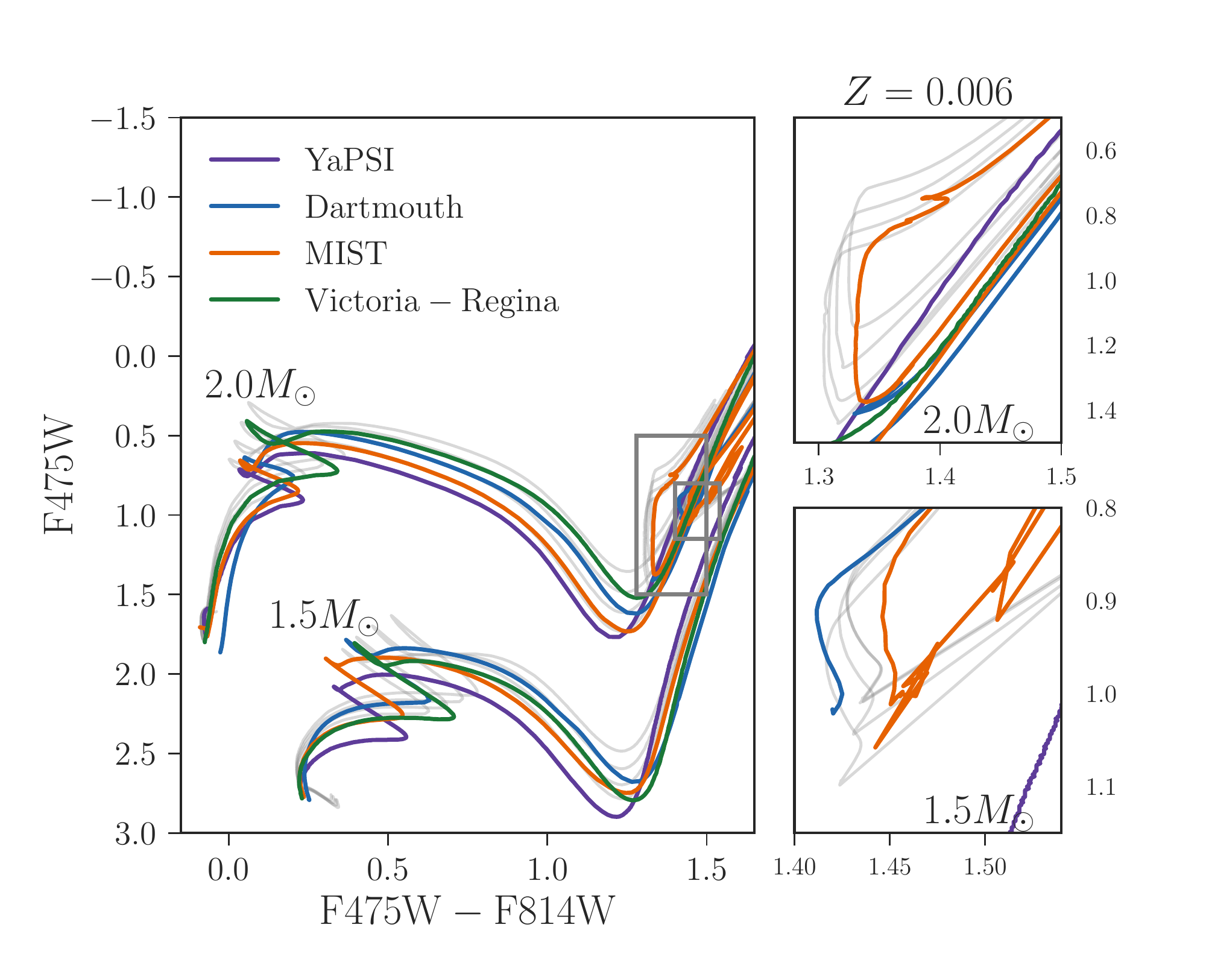}
\caption{Left: Hertzsprung-Russell diagram (left) and CMD (right) showing model stellar evolutionary tracks at two masses from different stellar modeling groups. In gray are the PARSEC tracks from Figure \ref{fig:COV_HRD}.}
\label{fig:COV_HRD_extra}
\end{figure*}

To illustrate a sample of the different predictions between stellar modeling groups, Figure \ref{fig:COV_HRD_extra} is based on Figure \ref{fig:COV_HRD} but with the PARSEC core overshooting grid in gray, and  \href{http://www.cadc-ccda.hia-iha.nrc-cnrc.gc.ca/community/VictoriaReginaModels/}{Victoria-Regina}, \href{http://vo.aip.de/yapsi/description_2016.html}{YaPSI}, \href{http://stellar.dartmouth.edu/\%7Emodels/fehp00afep0.html}{Dartmouth}, and \href{http://waps.cfa.harvard.edu/MIST/model_grids.html}{MIST} tracks over-plotted. For the CMD on the right panel, we applied the same bolometric corrections as we have for PARSEC \citep[see Section \ref{sec:effectcov} and][]{Girardi2008} for Victoria-Regina and YaPSI who publish isochrones in HST filter systems, but not tracks. Some specific differences between each modeling group are listed in Table \ref{tab:stelgroups}.

Even in this limited example, the predictions from one model to another nearly cover the entire HRD and CMD space of the PARSEC core overshooting grid (though the track with $\lambdac=0.5$ is much more likely to explain the data than the other overshooting values plotted). However, the predicted CMD morphologies are different enough that they may not be degenerate, especially considering any differences in predicted lifetimes (not shown) between modeling groups. With high signal-to-noise observations, a large number of stars, and filters chosen to maximize the separation of CMD features, one could distinguish between the different predicted CMD morphologies from each modeling group. 

The differences due to the careful decisions and their implementations between one stellar modeling group and another are model predictions that can and should be systematically tested against observations.  The statistical (Bayesian) treatment we have presented is applicable for exactly this purpose, as it is agnostic of stellar model and uncertain parameters. With a uniform binning scheme, only one further step would be necessary to compare one model to another, which is to calculate the model evidence.

\begin{deluxetable*}{lllll}
\tablecaption{Stellar Tracks from Five Modeling Groups}
\tablecolumns{5}
\tablewidth{0pt}
\tablehead{
\colhead{Source} &
\colhead{$Z_i$} &
\colhead{$Y_i$} &
\colhead{$\alpha_{\rm MLT}$} &
\colhead{Heavy Element Mixture}
}
\startdata
PARSEC V1.2S & 0.006 & 0.259 & 1.74 & \citet{Grevesse1998}, \citet{Caffau2011} \\
YaPSI	& 0.005416 & 0.25 & 1.91804 & \citet{Grevesse1998}  \\
Victoria-Regina	& 0.006 & 0.247 & 1.89 & \citet{Anders1989, Grevesse1990, Grevesse1991} \\
Dartmouth	& 0.006 & 0.254 & 1.938 & \citet{Grevesse1998}  \\
MIST	& 0.00582 & 0.2577 & 1.82 & \citet{Asplund2009}  \\
\enddata
\tablecomments{Stellar evolution track parameters of those shown in Figure \ref{fig:COV_HRD_extra}.}
\label{tab:stelgroups}
\end{deluxetable*}

\acknowledgements
 PR thanks Charlie Conroy, Benjamin Johnson, and Phillip Cargile for many helpful discussions. This material is based upon work supported by the National Science Foundation under Award No. 1501205. Support for this work was also provided by NASA through grant number AR-13901 from the Space Telescope Science Institute. All of the data presented in this paper were obtained from the Mikulski Archive for Space Telescopes (MAST). STScI is operated by the Association of Universities for Research in Astronomy, Inc., under NASA contract NAS5-26555. Support for MAST for non-HST data is provided by the NASA Office of Space Science via grant NNX09AF08G and by other grants and contracts. Computations in this paper were run on the Odyssey cluster supported by the FAS Division of Science, Research Computing Group at Harvard University.

This research has made use of NASA's Astrophysics Data System, NASA/IPAC Extragalactic Database (NED) which is operated by the Jet Propulsion Laboratory, California Institute of Technology, under contract with the National Aeronautics and Space Administration, the IPython package \citep{PER-GRA:2007}, Astropy, a community-developed core Python package for Astronomy \citep{2013A&A...558A..33A}, TOPCAT, an interactive graphical viewer and editor for tabular data \citep{2005ASPC..347...29T}, SciPy \citep{jones_scipy_2001}, and NumPy \citep{van2011numpy}.

\bibliography{clusters}

\end{document}